\documentclass[10pt,journal]{IEEEtran}
\usepackage{multirow}
\usepackage{xcolor}
\usepackage{soul, xcolor}
\setstcolor{red}
\usepackage[ruled,linesnumbered]{algorithm2e}
\usepackage{fixltx2e}
\usepackage[T1]{fontenc}
\usepackage{microtype}
\usepackage{booktabs}
\usepackage{tabularx}

\newcommand{\CLASSINPUTbaselinestretch}{1.0} 
\newcommand{\CLASSINPUTinnersidemargin}{1in} 
\newcommand{\CLASSINPUToutersidemargin}{1in} 
\newcommand{\CLASSINPUTtoptextmargin}{1in}   
\newcommand{\CLASSINPUTbottomtextmargin}{1in}

\usepackage{ifpdf}

\ifCLASSOPTIONcompsoc
  \usepackage[nocompress]{cite}
\else
  \usepackage{cite}
\fi
\ifCLASSINFOpdf
  \usepackage[pdftex]{graphicx}
\else
   \usepackage[dvips]{graphicx}
\fi
\usepackage{amsmath}
\usepackage{acronym}
\usepackage{algorithmic}

\usepackage{array}

\usepackage{mdwmath}
\usepackage{mdwtab}

\usepackage{eqparbox}

\ifCLASSOPTIONcompsoc
  \usepackage[caption=false,font=footnotesize,labelfont=sf,textfont=sf]{subfig}
\else
  \usepackage[caption=false,font=footnotesize]{subfig}
\fi
\usepackage{fixltx2e}

\ifCLASSOPTIONcaptionsoff
  \usepackage[nomarkers]{endfloat}
 \let\MYoriglatexcaption\caption
 \renewcommand{\caption}[2][\relax]{\MYoriglatexcaption[#2]{#2}}
\fi
\usepackage{url}

\ifCLASSINFOpdf
  \usepackage[pdftex]{thumbpdf}
\else
  \usepackage[dvips]{thumbpdf}
\fi
\begin{document}
%

\title{ System and Design Technology Co-optimization of SOT-MRAM for High-Performance AI Accelerator Memory System}

%
%
%
%

\author{Kaniz~Mishty~\IEEEmembership{}
        and~Mehdi~Sadi,~\IEEEmembership{Member,~IEEE,}
\IEEEcompsocitemizethanks{\IEEEcompsocthanksitem The authors are with the Department of Electrical and Computer Engineering, Auburn University, Auburn, AL 36849 USA \protect\\
E-mail: kzm0114@auburn.edu; mehdi.sadi@auburn.edu
}
\thanks{This work was supported in part by the National Science Foundation (NSF) under Grant Number CRII-2153394.}}

%
%

\markboth{IEEE Transactions on Computer-Aided Design,~Vol.~XX, No.~X, August~202X}%
{Shell \MakeLowercase{\textit{et al.}}: Bare Advanced Demo of IEEEtran.cls for IEEE Computer Society Journals}
%



\IEEEtitleabstractindextext{%
\begin{abstract}
System on Chips (SoCs) are now designed with their own AI accelerator segment to accommodate  the ever-increasing demand of Deep Learning (DL)  applications. With powerful Multiply and Accumulate (MAC) engines for matrix multiplications, these accelerators show high computing performance. However, because of limited memory resources (i.e., bandwidth and capacity), they fail to achieve optimum system performance during large batch training and inference. In this work, we propose a memory system with high on-chip capacity and bandwidth to shift the gear of AI accelerators from memory-bound to achieving system-level peak performance. We develop the memory system with Design Technology Co-optimization (DTCO)-enabled customized Spin Orbit Torque (SOT)-MRAM as large on-chip memory through System Technology Co-optimization (STCO) and detailed characterization of the DL workloads. 
Our workload-aware memory system achieves 8$\times$ energy and 9$\times$ latency improvement on Computer Vision (CV) benchmarks in training and 8$\times$ energy and 4.5$\times$ latency improvement on Natural Language Processing (NLP) benchmarks in training while consuming only around 50\% of SRAM area at iso-capacity. 
\end{abstract}

}

\maketitle

\IEEEdisplaynontitleabstractindextext

%
\IEEEpeerreviewmaketitle

\section{Introduction}
\label{intro}

\IEEEPARstart{T}{he} proliferation of Artificial Intelligence (AI) and Deep Learning (DL) has precipitated the computing hardware community to continually design innovative AI/DL accelerators with  large data processing capabilities. Research shows that the AI/DL model accuracy improves as training data set size grows \cite{DL_data1}. With increasing data set, model size also grows. Consequently, memory demand in AI/DL accelerators will also grow asymptotically linearly with model and data size \cite{DL_data1} \cite{dnn_survey}. As a result, the bottleneck for state-of-the-art AI/DL models in the accelerator hardware is now memory rather than data and compute availability, and we expect this trend to worsen in the future \cite{dnn_survey}\cite{tpu}\cite{amp100_gpu}.

The lack of efficient and high-performance data flow between the computing and memory element (i.e., the memory wall or memory bottleneck) masks the improvement coming from the efficient compute system \cite{cao2021mobile}. One promising solution to the memory bottleneck of AI-specific workload is to increase the on-chip memory capacity\cite{park2018deep}. For both training and inference, the on-chip memory capacity in the accelerator needs to be increased to ensure that the intermediate activations, as well as the weights of the current layer, can be loaded. Moreover, significantly more memory is required during training to store the gradients and optimizer states. Inadequate on-chip memory capacity causes frequent DRAM accesses which exacerbates energy costs, as well as stalls the compute cores of AI/DL accelerator until the data is fetched. Because of this large capacity demand, an SRAM-based on-chip memory system can be detrimental due to leakage energy and area inefficiency.

\begin{figure*}[ht]
	\centering
	\includegraphics[width=1.0\textwidth]{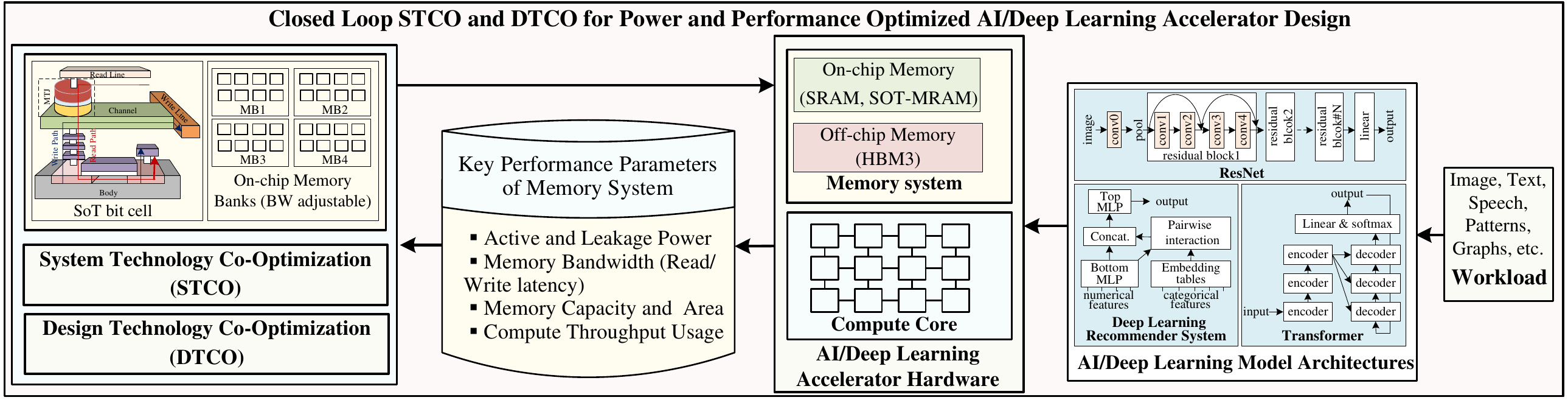}
	
	\caption{Workflow of closed-loop analysis for system and device level optimization for AI/Deep Learning Accelerator Design}
	\label{fig:paper_concept}
	
\end{figure*}

The promising features, such as high density, near-zero leakage power, immunity against radiation-induced soft errors, and CMOS compatibility of  emerging Spin-based non-volatile (NVM)  magnetic memory (i.e., MRAM)  technologies, attracted researchers from academia and industry \cite{memory_trend}. Spin Transfer Torque (STT) MRAM, has already shifted its gear from the R\&D phase to commercialization as the NAND-based embedded flash replacement \cite{dac_19} \cite{recent-progress-in-SOT_fab3}. However, MRAM in its regular form cannot be used in AI accelerators due to its slow write speed and high write energy \cite{ recent-progress-in-SOT_fab3}\cite{optimized_SOT_imec}.

STT-MRAM, a two-terminal magnetic memory with Magnetic Tunnel Junction (MTJ) as the storing element, flows a bidirectional spin-polarized current through the MTJ for read-write operation \cite{stt_eqn1}. The major challenges of STT-MRAM - poor write performance, Read Disturbance (RD), retention failure, \cite{recent-progress-in-SOT_fab3}\cite{dualport_fieldfree_fab2} - stem from two main reasons. First, the high write current flowing through the MTJ accounts for almost $10\times$ energy consumption as SRAM. Large write delay (> ns range) resulting from spin injection symmetry in switching the magnetic orientation of free layer belittles STT-MRAM's feasibility as an on-chip cache \cite{ultrafast_embedded_mem_fab4}. The stress on the dielectric oxide of the MTJ due to the large write current accelerates the time-dependent wear out of the cell \cite{tahoori_1}. Second, its shared read-write  path makes it vulnerable to RD.

SOT MRAM, considered the next generation of STT-MRAM, offers high performance without compromising reliability issues such as RD. SOT-MRAM is a three-terminal memory cell that uses MTJ as the storing element \cite{sot_model_kazemi}. By splitting the read-write path and using a different switching scheme, SOT-MRAM resolves all the challenges of STT-MRAM while retaining its every benefit \cite{recent-progress-in-SOT_fab3} \cite{dualport_fieldfree_fab2} \cite{ultrafast_embedded_mem_fab4} \cite{tahoori_1} \cite{size_dependent_switching_fab1}. Isolate read and write path allows the designer to optimize the read and write path independently, decreasing the write current and increasing the read-write operating margin, thus solving the RD-induced reliability issues. 
Though lacking mass-scale production from foundries due to early-stage manufacturing challenges, \cite{recent-progress-in-SOT_fab3} \cite{optimized_SOT_imec} 
\cite{dualport_fieldfree_fab2} \cite{ultrafast_embedded_mem_fab4} \cite{size_dependent_switching_fab1} \cite{sot_0.35ns_write} have demonstrated the successful fabrication of SOT-MRAM with attractive specifications. Its attractive features, such as high density, reliability and endurance, zero leakage, read-write latency comparable to SRAM, and research effort to enable mass production make it one of the best candidates for AI accelerator memory system where large on-chip memory is a must for training and inference.

The performance of an AI accelerator depends on both the compute and memory throughput of the device. While most accelerators have enough compute throughput, their performance is limited by memory throughput operating in the \emph{memory bound} region. To address the \emph{memory bound} problem of the AI hardware, in this paper, we perform a closed-loop STCO on AI workloads and DTCO on SOT-MRAM to present a hybrid memory system. To our knowledge, this is the first work that analyzes and evaluates the performance of SOT-MRAM as the on-chip memory of AI accelerators targeting both inference and training. The STCO-DTCO methodology is shown in Fig. \ref{fig:paper_concept}, and the key contributions of the paper are highlighted as follows.

\begin{itemize}
    \item We present a power and performance-optimized hybrid memory system for Deep Learning (DL) accelerators through a workload-aware STCO and DTCO. Comprised of off-chip HBM3 DRAM, on-chip SRAMs, and DTCO-enabled SOT-MRAM, the hybrid memory system can support the training and inference of DL workloads. We perform a closed-loop STCO and DTCO by taking into account the (i) System performance attributes (e.g., throughput and energy cost); (ii) Architectural and micro-architectural attributes (e.g., compute resources utilization, memory bandwidth) (iii) Workload attributes at both training and inference (e.g., runtime action counts, dataflow and data reuse) to reach the Pareto optimal solution.
    
    \item Using the Deep Learning models’ execution profiles,  DTCO enables device and circuit level customization of read/write bandwidth, retention time, and capacity of SOT-MRAM memory banks to meet the bandwidth and capacity demands of DL workloads. To achieve dynamic runtime optimization of the power and performance of the accelerator hardware for diverse workloads, memory banks are individually optimized with various bandwidths and capacities.

    \item Finally, using various DNN benchmarks,  we provide a comparative analysis of the existing SRAM-based memory system and the proposed DTCO-STCO optimized hybrid memory system for AI accelerators.
\end{itemize}

The rest of the article is organized as follows. Section \ref{background} discusses the background. In Section \ref{workload_profiling}, we present the analytical model for DNN workload profiling, followed by the DTCO of SOT-MRAM in Section \ref{dtco_MRAM}. Sections \ref{result_analysis} and \ref{rltd_work} present the results \& analysis, and related works, respectively, following the conclusion in Section \ref{conclusion}.

\section{Background}
\label{background}
\subsection{AI/DL Applications}

\begin{figure*}[ht]
    
    \centering
    \includegraphics[scale=0.8]{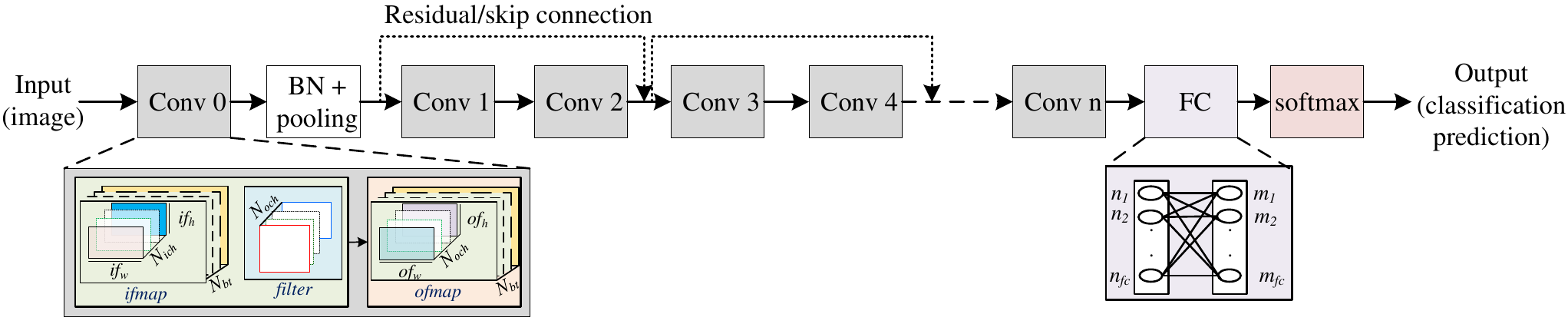}
    
    \caption{CV model (CNN/DNN) abstract architecture. Deep convolution (Conv) layers with residual/skip connection followed by fully connected (FC) layer/s. For symbol meaning please see Table \ref{sys_param}.}
    \label{fig:conv_op}
       
\end{figure*}
\subsubsection{Computer Vision (CV) and Pattern Recognition}

CV models, also called Convolutional/Deep Neural Networks (CNN/DNN),  are the stacks of convolution layers connected straight and/or through residual connection \cite{resnet} to extract the objects’ features, and a few Fully Connected (FC) layers at the end to classify the objects. Image classification,  captioning, reconstruction and object/instance segmentation are the scopes of CV models.  Deep Residual Networks, having convolutional layers at their core, dominate the CV domain. The input images are convolved with the filter weights to produce the output feature map (\emph{OFMAP}). The \emph{OFMAP} goes through the pooling and normalization layers to act as input (\emph{IFMAP}) to the next layer. The linear and softmax layer at the end finally recognizes the image (Fig. \ref{fig:conv_op}). The size of each data entity (IFMAP, OFMAP, and Weights) depend on the model architecture.

\subsubsection{Natural Language Processing (NLP)}
\begin{figure*}[ht]
    \centering
    \includegraphics[scale=0.65]{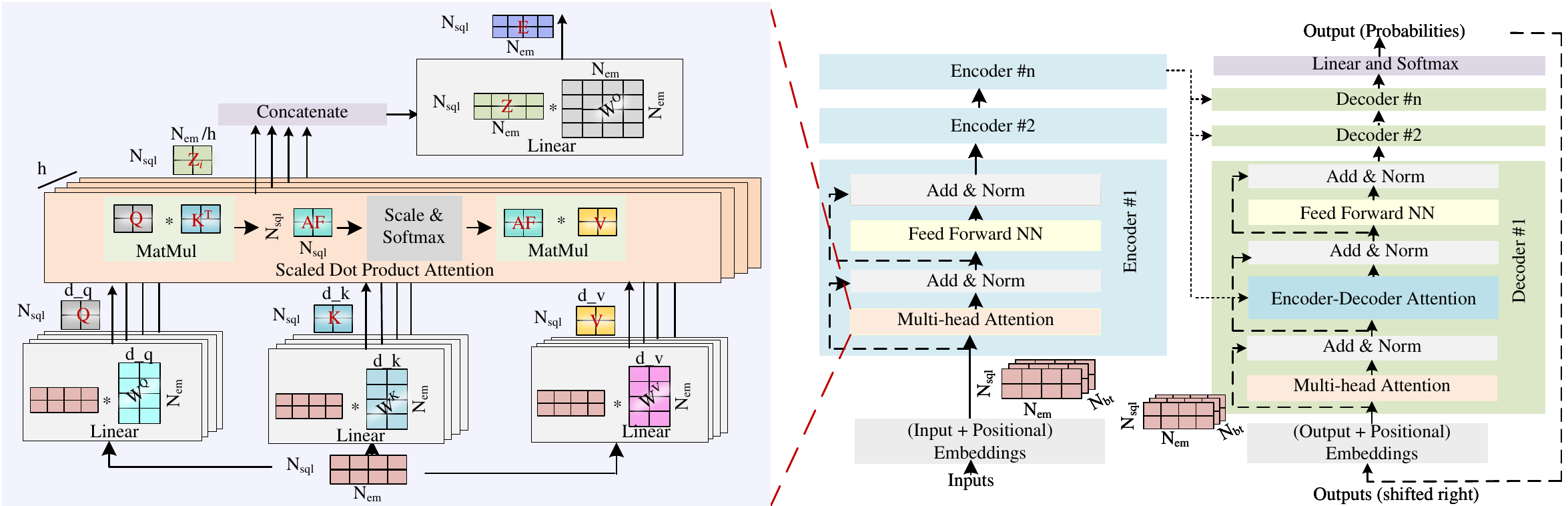}
    
    \caption{Transformer model workflow breakdown}
   
    \label{fig:transformer}
\end{figure*}

Language modeling deals with processing sequential data. Recurrent Neural Networks (RNN), Long Short Term Memory (LSTM), and Gated Recurrent Unit (GRU) have been used in language modeling until the state-of-the-art Transformer \cite{vaswani2017attention} model is introduced. NLP models are used in machine translation, text summarization, speech recognition, syntactic and semantic parsing, question answering, dialog system etc. In Transformer-based models \cite{vaswani2017attention}, the input sequence propagates through the embedding layer and different sublayers of the encoder stacks to extract different linguistic features and inter-token dependency of the input sequence. The decoder stacks then generate the output sequence by taking the encoded input sequence from the encoder stack and the output sequence generated by itself in the previous timesteps (Fig \ref{fig:transformer}). The input sequence multiplied by different layer weights takes different activation names and shapes throughout the model operation.

\subsection{AI/DL Accelerators}

At the core of AI/DLs is the matrix-matrix/vector multiplication (GEMM) with massive parallelism. Exploiting this parallelism,  Systolic Array (SA) based architecture \cite{tpu} have been used to accelerate the computations. Different dataflows, such as row stationary, output stationary, weight stationary, have been evolved to maximize the reuse and reduce the data movement. Off-chip DRAM access being 100-200 times more energy and latency expensive than any ALU operation or on-chip access \cite{eyeriss} plays a crucial role in determining the overall system performance. Another non-conventioanl type of architecture, In-Memory Computing (IMC) \cite{imc_survey} has recently  evolved to address the data communication cost for DNN accelerators. However, in this work, we focus on reducing the off-chip memory access for conventional DNN accelerator architectures \cite{tpu}, \cite{eyeriss}\cite{nvdla} by increasing the on-chip Global Buffer (GLB) size with SOT-MRAM. 

\begin{figure}[ht]
	\centering
	\includegraphics[scale=0.72]{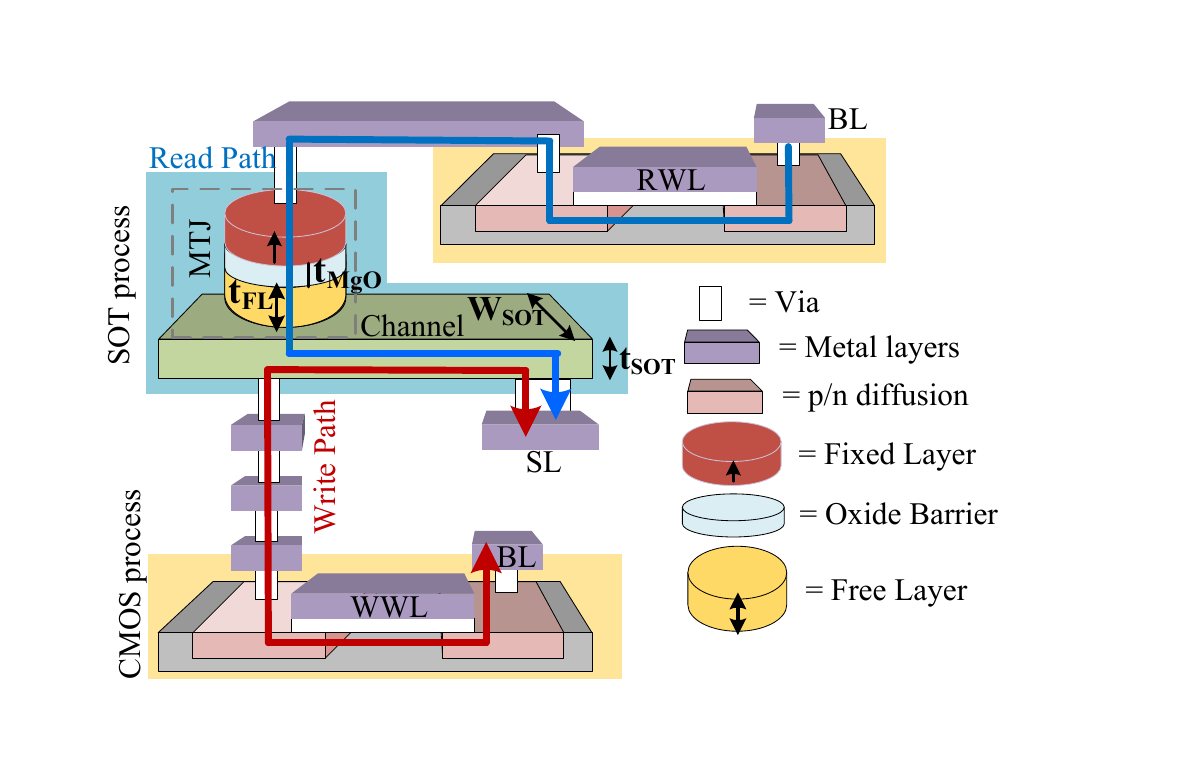}
	
	\caption{Physical structure of a SOT-MRAM bit cell highlighting separate read (along blue line) and write (along red line) path}
	\label{fig:sot_bitcell}	
	
\end{figure}

\subsection{SOT-MRAM}
\subsubsection{Physical Structure} 
With MTJ \cite{stt_eqn1} as storing element, the SOT-MRAM is a 
three terminal device. Depending on the type of bit cell, there are three to four lines to control the read-write operation. In this work, we consider a two transistor one SOT (2T1SOT) bit cell architecture that requires two access transistors, (i) \emph{Read Wordline (RWL)}, (ii) \emph{Write Wordline (WWL)}, (iii) \emph{Bit Line (BL)}, and (iv) \emph{source Line (SL)} to accommodate separate read-write access path \cite{sot_model_kazemi} \cite{2T1SOT} (Fig. \ref{fig:sot_bitcell}). The MTJ stack, with its free layer at the interface, is placed on top of a SOT layer (i.e., channel) to ensure SOT-induced switching. The SOT layer is composed of heavy metals or topological insulators \cite{manchon2019current}.

\subsubsection{Read-Write Operation}
Upon the activation of RWL, a small amount of current is passed through BL and grounded SL. The resistive state of the MTJ is captured by sensing the voltage across it and comparing the voltage with a reference value \cite{dualport_fieldfree_fab2}. Low resistive state ($R_{P}$) and high resistive state ($R_{AP}$) represents bit 0 and 1 respectively.
The write operation of MTJ-based MRAM involves switching the resistive status of MTJ. In SOT-MRAM, switching occurs due to Spin Orbit Torque (SOT) effect. Unlike STT-MRAM, a current is passed through the SOT layer to change the MTJ resistive state by switching the magnetic orientation of the free layer. 
A bidirectional write current flows through BL and SL during write operation. The potential of BL and SL changes depending on the bit value written in the cell. For example, to write `1', current flows from BL to SL and vice versa to write `0' \cite{dualport_fieldfree_fab2} \cite{tahoori_1}.

\section{DNN WORKLOAD PROFILING}
\label{workload_profiling}
Profiling the target workload is a prerequisite for designing an accelerator for the target workload. Assuming that we have a powerful computing system to handle the exhaustive computations of the DL workload, we focus on providing efficient data movement between the compute and memory system to ensure 100\% utilization of computing resources by introducing the workload-aware hybrid memory system. We propose the hybrid memory system by analyzing the Deep Learning model workloads from CV   and NLP domain.
We analytically model the on-chip bandwidth requirement and memory access patterns of different parts of the workload during inference and training, \emph {Memory and Compute Model}, to develop the memory system for TPU-like \cite{tpu} DNN accelerators.

\begin{figure}[ht]
    \centering
    \includegraphics[width=0.45\textwidth]{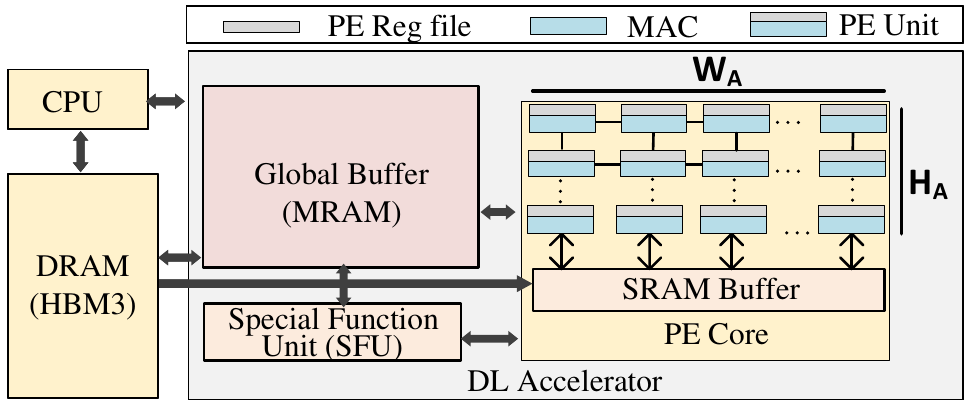}
    
    \caption{Block diagram of Accelerator architecture}
    \label{fig:system_architecture}
\end{figure}

\begin{figure*}[ht]
    \centering
    \includegraphics[scale=0.7]{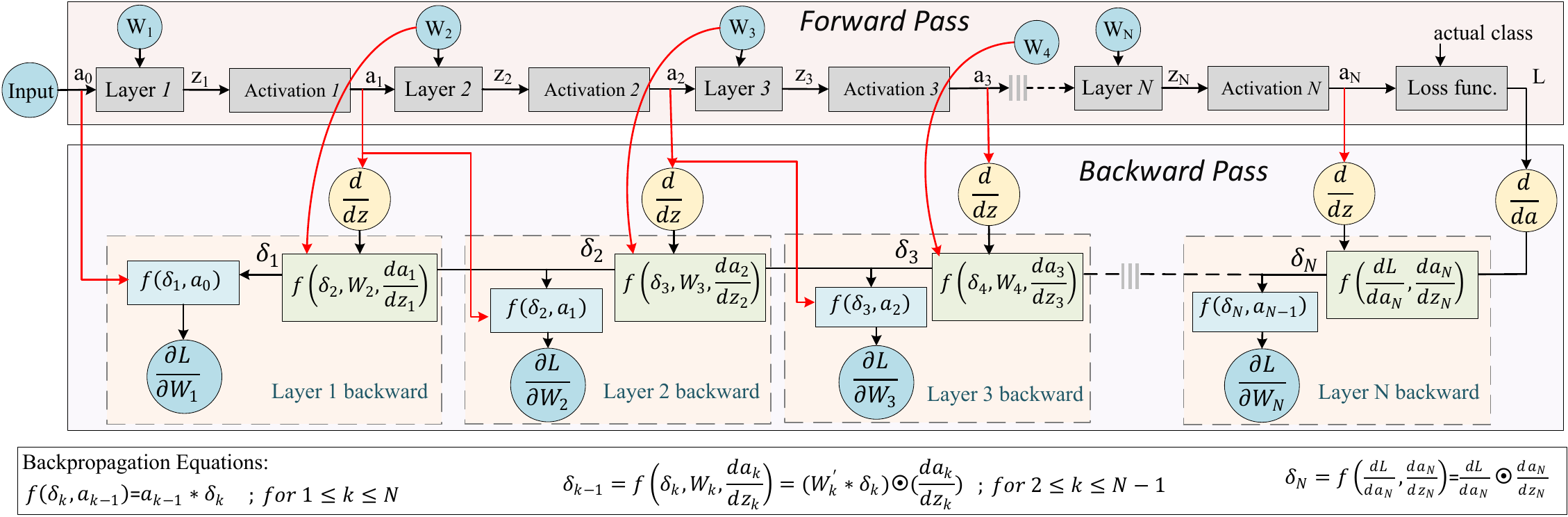}
    
    \caption{Computational graph of DNN training}
    
    \label{fig:comp_graph}
\end{figure*}

\subsection{Memory Bandwidth Expression}
We express the required bandwidth (BW) as a function of compute resources and workload. $BW$ (bytes/sec)  is defined as the rate at which data needs to be transferred to/from memory by a processor to utilize the computation resources of the processor fully. Mathematically,
\begin{align}
\label{req_mem_bw}
   BW= \frac{F_p}{OI}
\end{align}
Where $F_p$ = Theoretical peak performance  (ops/sec) = number of operations the accelerator performs per sec. The $F_{p}$ of a $H_{A} \times W_{A}$  Processing Element (PE) array (Fig. \ref{fig:system_architecture}):

\begin{align}
    \label{peak_fp}
        F_{p} = H_A*W_A*F_{acc}
\end{align}
$F_{acc}$ = Operating frequency of the accelerator. $OI$ = Operational Intensity of Workload (ops/byte) = number of operations performed per byte accessed. It is a measure of parallelism of the workload. In the subsequent subsections, we will formulate the $OI$ of Conv. and FC layer to find their BW, respectively. Note that the read and write bandwidth will not be the same for these workloads. 

\subsubsection{Read Bandwidth ($BW_{RD}$) of Conv. layer}

To formulate an expression for \emph{OI} of convolution workload: First, we determine the total number of MAC operations, $T_{MAC}$, performed by a $H_{A} \times W_{A}$ PE array per clock cycle
\begin{align}
    T_{MAC}\;=H_{A}\;*\;W_{A}
\end{align}
Second, we figure out how many bytes should be read from memory to utilize all PEs of the accelerator in one clock cycle. In a row stationary dataflow \cite{eyeriss}, it takes ($k_{h}*k_{w}+of_{h}*of_{w})*d_{w}$ bytes of data ($d_{w}$ = data type in bytes, i.e., FP32, BF16 etc.) and \#$(of_{h} * of_{w} * k_{h} *k_{w})$ PEs to generate the partial ofmaps corresponding to one input channel. Depending on the size of the PE array, in each iteration (one complete use of accelerator), multiple input channels can be fit. The input channels (i.e., no. of partial ofmaps) computed by the PE array in each iteration:

\begin{align}
    N_{ich\_per\_stp}\;=\frac{H_A*W_A}{of_{h}*of_{w}*k_h*k_w} 
\end{align}
Total bytes read from memory to utilize all PEs:
\begin{equation}
\begin{aligned}
    T_{byte}=\frac{H_A*W_A}{k_h*k_w*of_{h}*of_{w}}* 
    {(k_h*k_w+if_{h}*if_{w})*d_{w}}
\end{aligned}
\end{equation}
We divide the total number of MAC operations, $T_{MAC}$, by the total bytes accessed, $T_{byte}$, to find $OI$:
\begin{equation}
    \begin{aligned}
    \label{oi}
        OI=\frac{k_h*k_w*of_{h}*of_{w}}{d_{w}*(k_h*k_w+if_{h}*if_{w})}
    \end{aligned}
\end{equation}
Substituting the expression of $OI$ in equation  \eqref{req_mem_bw} gives $BW_{RD}$ as a function of array size and workload:
\begin{equation}
    \begin{aligned}
        BW_{RD} & = \frac{(k_h*k_w\;+if_{h}*if_{w})*d_{w}}{k_w*k_h*of_{h}*of_{w}}* H_{A}*W_{A}*F_{acc}
    \end{aligned}
\end{equation}
For the symbol meanings, please see Fig. \ref{fig:conv_op} and Table \ref{sys_param}.

\subsubsection{Write Bandwidth ($BW_{WR}$) of Conv. Layer}
Partial ofmap of a single input channel requires \#($of_{h}*of_{w}*k_{h}*k_{w}$) PEs. Therefore, $H_{A}\times W_{A}$ PEs generate $(H_{A}*W_{A})/(of_{h}*of_{w}*k_{h}*k_{w})$ ofmaps in each iteration. Each partial ofmap contains $of_{h}*of_{w}$ elements. The total output bytes generated by the PE array in one iteration is, equivalently, the write bandwidth:

\begin{table}[]
\centering
\caption{CNN and systolic array parameters nomenclature}
\label{sys_param}
\begin{tabular}{|l|l|}
\hline
$W_{A}, H_{A}$   & Accelerator array width \& height (PEs) \\ \hline
$k_{w}, k_{h}$   & Kernel width \& height \\ \hline
$of_{w}, of_{h}$  & Output feature map width \& height \\ \hline
$if_{w}, if_{h}$   & Input feature map width \& height \\ \hline
$N_{ich}, N_{och}$ & No. of input \& output channel \\ \hline
$N_{bt}$  & Batch size \\ \hline
$n_{fc}, m_{fc}$  & No. of neurons in input \& output FC layer  \\ \hline
\end{tabular}
\end{table}

\begin{equation}
\begin{aligned}
    \label{wr_bw}
    BW_{WR} = \frac{H_{A}*W_{A} * F_{acc} * d_{w}}{k_{h}*k_{w}}\;
\end{aligned}
\end{equation}

\begin{table}[ht]
\setlength{\tabcolsep}{3pt} 
\centering
\caption{RD/WR bandwidth expression of FC layer for different cases}
\label{table:bw_expression}
\begin{tabular}{|cc|c|c|}
\hline
\multicolumn{2}{|c|}{Cases}& $BW_{RD}$& $BW_{WR}$\\ \hline
\multicolumn{1}{|c|}{\multirow{4}{*}{$M<H_{A}$; $N<W_{A}$}}& \multirow{2}{*}{$K<W_{A}$}& \multirow{2}{*}{$\frac{M*N+K*M}{N+K}$}& \multirow{2}{*}{$\frac{K*N}{2*N+K-1}$}\\
\multicolumn{1}{|c|}{}&&&\\ \cline{2-4} 
\multicolumn{1}{|c|}{}& \multirow{2}{*}{$K\ge W_{A}$} & \multirow{2}{*}{$\frac{M*N+W_{A}*M}{N+W_{A}}$}& \multirow{2}{*}{$\frac{W_{A}*N}{2*N+K-1}$}\\
\multicolumn{1}{|c|}{}&&&\\ \hline
\multicolumn{1}{|c|}{\multirow{4}{*}{$M<H_{A}; N\ge W_{A}$}}  & \multirow{2}{*}{$K<W_{A}$}& \multirow{2}{*}{$\frac{M*W_{A}+K*M}{N+K}$}& \multirow{2}{*}{$\frac{K*W_{A}}{2*W_{A}+K-1}$}\\
\multicolumn{1}{|c|}{}&&&\\ \cline{2-4} 
\multicolumn{1}{|c|}{}& \multirow{2}{*}{$K\ge W_{A}$} & \multirow{2}{*}{$\frac{M*W_{A}+W_{A}*M}{2*W_{A}}$}& \multirow{2}{*}{$\frac{{W_{A}}^{2}}{2*W_{A}+K-1}$} \\
\multicolumn{1}{|c|}{}&&&\\ \hline
\multicolumn{1}{|c|}{\multirow{4}{*}{$M\ge H_{A}; N<W_{A}$}}  & \multirow{2}{*}{$K<W_{A}$}& \multirow{2}{*}{$\frac{H_{A}*N+K*H_{A}}{N+K}$}& \multirow{2}{*}{$\frac{K*N}{2*N+K-1}$}\\
\multicolumn{1}{|c|}{}&&&\\ \cline{2-4} 
\multicolumn{1}{|c|}{}& \multirow{2}{*}{$K\ge W_{A}$} & \multirow{2}{*}{$\frac{H_{A}*N+W_{A}*H_{A}}{W_{A}+N}$}& \multirow{2}{*}{$\frac{W_{A}*N}{2*N+K-1}$}\\
\multicolumn{1}{|c|}{}&&&\\ \hline
\multicolumn{1}{|c|}{\multirow{4}{*}{$M\ge H_{A};N\ge W_{A}$}} & \multirow{2}{*}{$K<W_{A}$}    & \multirow{2}{*}{$\frac{H_{A}*W_{A}+W_{A}*H_{A}}{W_{A}+K}$} & \multirow{2}{*}{$\frac{W_{A}*N}{2*N+K-1}$}\\
\multicolumn{1}{|c|}{}&&&\\ \cline{2-4} 
\multicolumn{1}{|c|}{}& \multirow{2}{*}{$K\ge W_{A}$} & \multirow{2}{*}{$\frac{H_{A}*W_{A}+W_{A}*H_{A}}{2*W_{A}}$} & \multirow{2}{*}{$\frac{{W_{A}^2}}{2*W_{A}+K-1}$}   \\
\multicolumn{1}{|c|}{}&&&\\ \hline
\end{tabular}
\end{table}

\begin{table}[]
\centering
\caption{Parameter nomenclature for Algorithm \ref{alg:dram_acc_cnt_infr} and \ref{alg:dram_acc_cnt_trng}}

\begin{tabular}{|l|l|}
\hline
$I, O, W$   & $ifmap,\; ofmap,\;weight$ size in MB \\ \hline
$RD_{DRAM}$   & DRAM Read access counts\\ \hline
$WR_{DRAM}$   & DRAM Write access counts \\ \hline
$RD_{GLB}$   & GLB Read access counts \\ \hline
$WR_{GLB}$   & GLB Write access counts \\ \hline
$GI, GO, GW$ & $ifmap, ofmap, weight$ Gradient size in MB\\ \hline
$mbpa$  & MB of data fetched per memory access \\ \hline
$layer\_f$   & Layer size (MB) combining $ifmap$, $ofmap$ \& \\   & $weights$ \\ \hline
$layer\_b$   & Layer size (MB) in backprop combining upstream,\\  
&  $ofmap$ \& $weight$ gradient  \\ \hline
$cum\;layer$ & Cummulative size of layer \\ \hline
$rd\_f, rd\_b$ & DRAM read access during forward \& backward pass \\ \hline
$wr\_f, wr\_b$  & DRAM write access during forward \& backward pass  \\ \hline
\end{tabular}
\label{alg_param}
\end{table}

\RestyleAlgo{ruled}
\SetKwComment{Comment}{/* }{ */}

\begin{algorithm}[hbt!]
\footnotesize
\caption{DRAM \& GLB access count at Inference }
\label{alg:dram_acc_cnt_infr}

\For{$i=1$ \KwTo $no.\;of\;layers$}
{$RD_{GLB} \gets \frac{I_{i}}{mbpa_{GLB}}$ \\

\eIf{$i=1$}
{$WR_{GLB} \gets \frac{I_{i}+O_{i}} {mbpa_{GLB}}$ \\
        \eIf{($I_{i}+W_{i})\leq GLB $} {$RD_{DRAM} \gets \frac{I_{i}+W_{i}}{mbpa_{DRAM}}$}{$RD_{DRAM} \gets \frac{I_{i}+W_{i}}{mbpa_{DRAM}}+ \frac{I_{i}+W_{i}-GLB}{mbpa_{DRAM}}$}}
        {$WR_{GLB} \gets \frac{O_{i}}{mbpa_{GLB}}$ \\
        \eIf{$O_{i-1} \leq UB$}
            {\eIf{$W_{i} \leq GLB $}{$RD_{DRAM} \gets \frac{W_{i}}{mbpa_{DRAM}}$}
                {$RD_{DRAM} \gets \frac{W_{i}}{mbpa_{DRAM}} + \frac{W_{i}-GLB}{mbpa_{DRAM}}$}}
            {$RD_{DRAM} \gets  \frac{I_{i}+W_{i}}{mbpa_{DRAM}}+\frac{(I_{i}+W_{i})-GLB}{mbpa_{DRAM}}$}}
    \eIf{$i=no.\;of\;layers$}{$WR_{DRAM} \gets \frac{O_{i}}{mbpa_{DRAM}}$}
    {\eIf{$O_{i}>GLB$}{$WR_{DRAM} \gets \frac{O_{i}-UB}{mbpa_{DRAM}}$}{$WR_{DRAM} \gets 0$}}
}

\end{algorithm}

\begin{algorithm}[hbt!]
\footnotesize
\caption{DRAM \& GLB access count at Training} 
\label{alg:dram_acc_cnt_trng}
$cum\;layer \gets 0$\;
$tmp \gets 0$\;

\For{$i=1$ \KwTo $no.\;of\;layers$}
{   $layer\_f_{i} \gets I_{i} + O_{i} + W_{i}$ \;
    $layer\_b_{i} \gets GI_{i} + GO_{i} + GW_{i}$ \;
    $layer\;(i) \gets layer\_f_{i} + layer\_b_{i}$ \;
    $cum\;layer(i) \gets tmp + layer\;(i)$ \;
    $tmp \gets cum\;layer(i)$ \;

$RD_{GLB} \gets \frac{3*I_{i}+O_{i}+5*W_{i}}{mbpa_{GLB}}$ \\
$WR_{GLB} \gets \frac{2*I_{i}+2*O_{i}+3*W_{i}}{mbpa_{GLB}}$ \\ 
\eIf{$cum\;layer(i) \leq GLB$}
{\If{$i=1$}{$rd\_f(i) \gets \frac{I_{i}+W_{i}}{mbpa_{DRAM}}$}
\If{$i=no.\;of\;layers$}{$wr\_f(i) \gets \frac{O_{i}}{mbpa_{DRAM}}$ }
$rd\_f(i) \gets \frac{W_{i}}{mbpa_{DRAM}}$ \;
$rd\_b(i) \gets 0$ \;
$wr\_f(i) \gets 0$ \;
}
{
\eIf{($i\neq 1$) AND ($O_{i-1} \leq GLB$)}{$rd\_f(i) \gets \frac{W_i}{mbpa_{DRAM}}$}
{\eIf{$I_{i}+W_{i} \leq GLB$}{$rd\_f(i) \gets \frac{I_{i}+W_{i}}{mbpa_{DRAM}}$}{$rd\_f(i) \gets \frac{I_i+W_i}{mbpa_{DRAM}}+\frac{I_i+W_i-GLB}{mbpa_{DRAM}}$}}

\eIf{($GI_i+GO_i+GW_i \leq GLB$)}
{$wr\_f(i) \gets 0$ \\
$rd\_b(i) \gets 0$}
{$wr\_f(i) \gets \frac{GI_i + GO_i + GW_i}{mbpa_{DRAM}} $ \\
$rd\_b(i) \gets \frac{GI_i + GO_i + GW_i}{mbpa_{DRAM}}$}
}
$wr\_b(i) \gets \frac{W_{i}}{mb\_per\_acs}$
}

\end{algorithm}

\subsubsection{$BW_{RD}$ \& $BW_{WR}$ of FC layer}
\label{bw_fc_layer}
The systolic array is a widely used architecture to perform GEMM operation \cite{tpu}. 
Depending on the array dimension ($H_{A}\times W_{A}$) and operand matrix dimension (input matrix: $K\times M$, weight matrix: $M\times N$, and output matrix: $K\times N$), we formulate required Read and Write GLB bandwidth for four different cases: (i) Weight matrix dimensions (both) are less than the systolic array dimensions ($M<H_{A}, N<W_{A}$), (ii) Height of weight matrix is less than the height of systolic array, but the width of weight matrix is larger than or equal to the width of the systolic array ($M<H_{A}, N \ge W_{A}$), (iii) Height of weight matrix is larger than or equal to the height of systolic array, but width of the weight matrix is less than the width of the systolic array  ($M \ge H_{A}, N<W_{A}$), and (iv) Both height and width of weight matrix are larger than or equal to the height and width of systolic array respectively ($M \ge H_{A}, N\ge W_{A}$).


In a weight stationary dataflow, it takes $N$ clock cycles to load the weight matrix into the systolic array. Once the weights are loaded, the input matrix is streamed from left to right and the outputs are collected downward. 
The input matrix's first column reaches the weight matrix's last column at $2N$ clock cycles. The last (or $K^{th}$) column of the input matrix reaches the last column of weight matrix after $2N+K-1$ clock cycles and generates the output matrix, $K\times N$. 
Based on the above dataflow and mapping, the peak read-write bandwidth per clock cycle for different cases is summarized in Table \ref{table:bw_expression}. The expressions are shown for weight stationary dataflow. 

From the transformer-based NLP model architecture (Fig. \ref{fig:transformer}), we observe that the dominant operations are the GEMM operations. As a result, we model the read-write bandwidth requirement for different layers of the transformer-based model same as the read-write bandwidth of FC layer. Another dominant operation after GEMM is softmax operation, performed mostly on the scaled attention filter matrix, AF of size $N_{sql} \times N_{sql}$. $\sigma(AF)ij=\frac{e^{AF_{ij}}}{\sum_{i=1}^{N_{sql}}e^{AF_{ij}}}$. The softmax operation is generally performed in the Special Function Unit (SFU)\cite{amp100_gpu} (Fig. \ref{fig:system_architecture}). The bandwidth requirement of the softmax operation depends on the hardware architecture, mapping, and $AF$ matrix dimension. Assuming that the SFU contains $1 \times H_{A}$ units, each capable of performing one exponential operation, followed by an accumulator for accumulating the exponentials, and a regular ALU for performing the division the bandwidth of softmax operation on SFU is estimated as $BW_{softmax} = d_{w} * H_{A}$. 

\subsection{Memory Access Patterns}
\label{mem_access}
Our proposed memory system consists of HMB3 (off-chip DRAM memory), a large GLB with multiple SOT-MRAM banks, a smaller double-buffered SRAM, and PE reg file specific to each PE unit (Fig. \ref{fig:system_architecture}). The banks inside SOT-MRAM are optimized through a DTCO between the SOT-MRAM parameters and the workload requirements. The double-buffered SRAM holds the weights and partial outputs. In this subsection, we analyze the memory access patterns of CV and NLP models for the proposed memory system.


The required number of main memory accesses depends on the GLB size, weight, activation size, and dataflow. Assuming a fixed dataflow, weight stationary in this case, we model the memory access counts during inference and training as a function of the model's workloads and the GLB size in Algorithm \ref{alg:dram_acc_cnt_infr}, and \ref{alg:dram_acc_cnt_trng} for inference and training, respectively. During inference, inputs (e.g., images, tokens) are read from HBM3, written to GLB, and read from GLB to be operated inside PEs core. The read-only weights are directly loaded from HBM3 to the register file of each PE unit, bypassing the GLB. Using   double-buffered SRAM, while the array is computing with loaded weights, the next set of weights is temporarily written to the SRAM buffer to hide the off-chip access latency behind the PE array computation latency. 
Suppose the GLB size is large enough to hold all samples in the minibatch. In that case, the data entity can be read all at once, resulting in the memory accesses equal to the algorithmic minimum memory accesses. Algorithmic minimum memory access represents the number of elements in the data entity \cite{parashar2019timeloop}. For weight gradient calculation, during backpropagation of the training, the inputs are read from GLB to PE core, assuming that the GLB is large enough to hold the input images along with the generated ofmap of the current layer, thus avoiding the DRAM accesses during backward pass. In convolution, the inputs can be reused multiple times for convolutional and filter reuse. It can also be reused multiple times during backpropagation to calculate the gradients of different filters. In Transformer-based NLP models, the embedded input can be reused thrice as input to Key, Query, and Value linear layer. It can also be reused thrice during backpropagation to calculate the weight gradient of the Key, Query, and Value linear layer. The training workflow is complicated and requires many more memory accesses (both off-chip and on-chip) compared to inference. For example, to calculate the weight gradients of Layer 1, it requires the current layer's activation gradient $\frac{da_1}{dz_1}$, input ($a_0$), next layer's weight ($W_{2}$) and the upstream gradient from Layer 2 ($\delta_{1}$) (Fig. \ref{fig:comp_graph}).

The pseudo code of Alg. \ref{alg:dram_acc_cnt_infr} models the inference memory access patterns. The inputs and weights must be loaded from DRAM for the first layer. Depending on the combined size of input \& weight matrix size, and GLB size, it requires either algorithmic minimum read accesses or more than that (lines 3-9 of Alg. \ref{alg:dram_acc_cnt_infr}). For the rest of the layers, if the $ofmap$ of the previous layer can fit in GLB, then no read accesses are required for input activation, as the $ofmap$ of the previous layer will act as the $ifmap$ to the next layer. Only the weights are read from DRAM for such layers (lines 12-20 of Alg. \ref{alg:dram_acc_cnt_infr}). The opposite case applies to write accesses: the $ofmap$ of the last layer must be written to the DRAM. For other layers, it needs to be written to DRAM depending on its size and GLB size (lines 22-30 of Alg. \ref{alg:dram_acc_cnt_infr}). No write accesses are required for weight matrices during inference. As the weights bypass the GLB during inference, the GLB read accesses are calculated from the $ifmap$ size for each layer (line 2). The write accesses are calculated from the $ofmap$ except for the 1st layer (line 11, 4). See Table \ref{alg_param} for symbol meanings.

The pseudo code of Algorithm \ref{alg:dram_acc_cnt_trng} models the training behavior. We initialize several temporary variables: $layer\_f_{i}$ (comprising of $ifmap$, $ofmap$, and $weight$ matrices of $i^{th}$ layer), $layer\_b_{i}$ (comprising of upstream gradient, $ofmap$, and $weight$ matrix gradients of $i^{th}$ layer), and $layer_{i}$ (combining  $layer\_f_{i}$ and $layer\_b_{i}$)  as shown in lines 4-6 in  Alg. \ref{alg:dram_acc_cnt_trng}. $cum\;layer(i)$ contains all layers' all entities up to $i^{th}$ layer (line 7-8, Alg. \ref{alg:dram_acc_cnt_trng}). If the GLB is large enough to hold $cum\;layer(i)$, we just need to read the $ifmap$ of the first layer \& $weight$ of all layers from DRAM during the forward pass and write all layers' updated $weight$ during the backward pass and last layer's $ofmap$ to DRAM during the forward pass (lines 11-20 \& 39, Alg. \ref{alg:dram_acc_cnt_trng}). Otherwise, the forward pass is the same as the inference (lines 22-30, Alg. \ref{alg:dram_acc_cnt_trng}). During the backward pass, depending on the size of upstream gradients, $ofmap$, and $weight$ gradients, it accesses the gradients from DRAM (lines 31-37, Alg. \ref{alg:dram_acc_cnt_trng}). The GLB read-write accesses are shown in lines 9-10 in Alg. \ref{alg:dram_acc_cnt_trng}. The $ifmap$ of each layer needs to be read twice, once during the forward pass and once during the backward pass. The upstream gradient, equal in size as $ifmap$, must be read once during the backward pass. The $ofmap$ is read once during the backward pass to calculate the upstream gradient. The weight is read 5 times (once during the forward pass, 4 times during the backward pass). The $ifmap$ and $ofmap$ are written twice, once during the forward pass and once during the backward pass. The $weight$ is written thrice, twice during forward pass and once during backward pass.

\section{DTCO of SOT-MRAM}
\label{dtco_MRAM}
To ensure  overall system performance for AI workloads, the memory system should have large on-chip memory to avoid frequent DRAM accesses, and the on-chip memory should have high bandwidth to prevent the system from being memory-bound while being energy efficient. In this section, we perform a DTCO of SOT-MRAM in bit-cell level based on the workload profiling done in section \ref{workload_profiling}.

\subsection{Optimizing critical switching current $I_{c}$}
In SOT-MRAM, the magnetic orientation of the free layer is switched by Spin-Orbit Torque induced by spin Hall and interfacial effects between the channel (i.e., SOT layer) and free layer (FL) of MTJ. An in-plane charge current is flown through the channel to generate a spin current that exerts a spin torque on the free layer, which rotates the free layer's magnetic orientation. The critical current density required to switch the magnetic orientation of FL is expressed as \cite{lee2013threshold}

\begin{equation}
    j_{c} = \frac{2e\mu_0 M_{s,FL}t_{FL}}{\hbar\theta_{SH}} (\frac{H_{k,eff}}{2} -\frac{H_x}{\sqrt{2}})
    \label{Ic3}
\end{equation}
Where $H_{k,eff}$ is the effective anisotropy field, $H_{x}$ is the applied field, $M_{s,FL}$ is the saturation magnetization of free layer, and $t_{FL}$ is its thickness. Our interest is in lowering the switching current to achieve low write energy.
Here, the free layer thickness $t_{FL}$ and spin Hall efficiency $\theta_{SH}$ act as a control knob for critical switching current. $\theta_{SH}$ is a material-specific parameter and its higher value is expected to reduce the switching current. The typical value of $\theta_{SH}$ in heavy metal alloys ranges between 0.1 to 0.5 \cite{manchon2019current}. However, recent topological insulators as SOT layer can have a very large $\theta_{SH}$. \cite{khang2018conductive} demonstrated $\theta_{SH} = 152$ with $BiSb$ thin films.

\subsection{Optimizing read-write pulse width}
\subsubsection{Read pulse width}
The reading of SOT-MRAM involves sensing the resistance of the MTJ. A small amount of current is passed through the MTJ stack and the voltage across the stack $V_{data+}$ or $V_{data-}$ is compared against a reference voltage $V_{ref} = \frac{1}{2}(V_{data+}+ V_{data-})$ to read out the stored bit. The read Sensing Margin $SM = |V_{ref} - V_{data}|$ is typically very small. Sensing and amplifying this small difference requires a strong and complex Sense Amplifier  that contributes to most of the read latency and energy. The SM is determined by the Tunnel Magneto Resistance ratio ($TMR\;ratio = \frac{R_{AP}-R_{P}}{R_{P}}$) of MTJ. A higher TMR ratio produces a larger SM by making $V_{data+}$ higher and $V_{data-}$ lower. Thus the TMR ratio is inversely proportional to the read latency\cite{3T-2SOT}. The higher the TMR window, the higher the read speed and the less effort required on the periphery. The typical range of the TMR ratio is between $100$ to $300\%$. The TMR is tunable by oxide thickness \cite{tsunekawa2005giant} as shown in Fig. \ref{fig:tmr_vs_oxide_rd_ltncy} (a). In SOT-MRAM, we can increase the oxide thickness, thanks to the decoupled read-write path of SOT-MRAM, to achieve a high TMR and increase the read speed without worrying about the large incubation time \cite{wang2013low}. 

\begin{table}[ht]
\setlength{\tabcolsep}{3pt}
\centering
\caption{DTCO control parameters \& their impact on Power, Performance and  Area (PPA)}
\label{dtco_table}
\begin{tabular}{|l|l|}
\hline
\textbf{DTCO Parameters}               & \textbf{Impact on PPA}                                                                             \\ \hline
Spin Hall angle $\theta_{SH}$ & $\theta_{SH} \uparrow$, $j_{c} \downarrow$,  Switching energy $\downarrow$                \\ \hline
Free layer thickness $t_{FL}$ & $t_{FL} \downarrow$, $j_{c} \downarrow$, Switching energy $\downarrow$, Area $\downarrow$           \\ \hline
\multirow{2}{*}{\begin{tabular}{@{}l@{}}{SOT layer dimension} \\ {$A_{SOT}$}\end{tabular}} & 
\multirow{2}{*}{\begin{tabular}{@{}l@{}}{$A_{SOT} \downarrow$, $\tau_{p} \downarrow$, Area $\downarrow$, Write Bandwidth $\uparrow$}\end{tabular}}\\
 & \\ \hline
Oxide thickness $t_{MgO}$ & $t_{MgO} \uparrow$, TMR $ \uparrow$, Read Bandwidth $ \uparrow$\\ \hline
\end{tabular}
\end{table}

\subsubsection{Write pulse width $\tau_{p}$}
The width of the write current pulse for switching is inversely proportional to the magnitude of the applied current density in the SOT layer $j_{sw}$ \cite{manchon2019current}
\begin{equation}
    \tau_{p} \propto \frac{1}{j_{sw}}
\label{tau}
\end{equation}

As the area of the SOT layer ($A_{SOT}$) is scaled down, the effective current density increases,  $j_{sw} \propto 1/(A_{SOT})$.  Successful switching should take place when $j_{sw} > j_{c}$. We can increase $j_{sw}$ by reducing the SOT layer dimension and decrease $j_{c}$ by increasing $\theta_{SH}$ or by decreasing $t_{FL}$. Thus we can achieve successful switching in much shorter pulse width (equation \ref{tau}). \cite{garello2014ultrafast} demonstrated the switching at 180ps, \cite{wu2021voltage} at 400ps, and \cite{garello2018sot} at 210ps. Switching in shorter pulse width ensures larger write bandwidth which is essential for memory systems used in AI/Deep Learning hardware. The key DTCO parameters of SOT-MRAM and their impact on Power, Performance and  Area (PPA) are listed in Table \ref{dtco_table}.

\section{Results and Analysis}
\label{result_analysis}

In this section, we provide the results and analysis of the STCO on the CV and NLP workloads during inference and training and present the optimum Power, Performance, and Area results by performing the DTCO of SOT-MRAM. We developed a MATLAB-based framework to implement our analytical \emph {Memory and Compute Model} to capture the relationship between the memory access counts and the memory hierarchy sizes in typical systolic array based AI accelerators. 
Unlike ScaleSim \cite{samajdar2020systematic} and Timelooop \cite{parashar2019timeloop} simulator, which only support profiling DNN workloads in inference mode to date, our model captures both training and inference behavior of CV and NLP models. We also verified our model's results with Timeloop in inference mode.

\subsection{Bandwidth Demand}
\label{bw_demand}


\begin{table*}[]
\centering
\caption{Parameters of NLP models}
\label{nlp_param}

\begin{tabular}{|c|c|c|c|c|c|c|c|}
\hline
Model & Enc. layer & Dec. layer & Attn. head & \begin{tabular}[c]{@{}c@{}}Word Embedding\\ ($N_{em}$)\end{tabular} & \begin{tabular}[c]{@{}c@{}}Intermediate dimension\\ ($d_{ff}$)\end{tabular} & \begin{tabular}[c]{@{}c@{}}Seq. length \\ ($N_{sql}$)\end{tabular} & \begin{tabular}[c]{@{}c@{}}Vocab. size\\ ($N_{vocab}$)\end{tabular} \\ \hline
Transformer  & 12 & 6 & 8 & 512 & 2048 & 1024 & 37000 \\ \hline
BERT & 12 & - & 12 & 768 & 3072 & 512 & 30522 \\ \hline
Distil BERT  & 6 & - & 12 & 768 & 3072 & 512 & 30522 \\ \hline
Mobile BERT  & 24 & - & 4 & 128 & 512 & 512 & 30522 \\ \hline
Squeeze BERT & 12 & - & 12 & 768 & 3072 & 512 & 30522 \\ \hline
Visual BERT  & 12 & - & 12 & 512 & 3072 & 512 & 30522 \\ \hline
GPT & - & 12 & 12 & 768 & 2048 & 512 & 40478 \\ \hline
GPT-2 & - & 12 & 12 & 768 & 2048 & 1024 & 50257 \\ \hline
GPT-3 & - & 96 & 96 & 12288 & 49152 & 2048 & 50257 \\ \hline
GPT-Neo & - & 24 & 16 & 2048 & 8192 & 2048 & 50257 \\ \hline
GPT-J & - & 28 & 16 & 4096 & 16384 & 2048 & 50400 \\ \hline
\end{tabular}
\end{table*}

\begin{figure}[ht]
    \centering
    \includegraphics[scale=0.6]{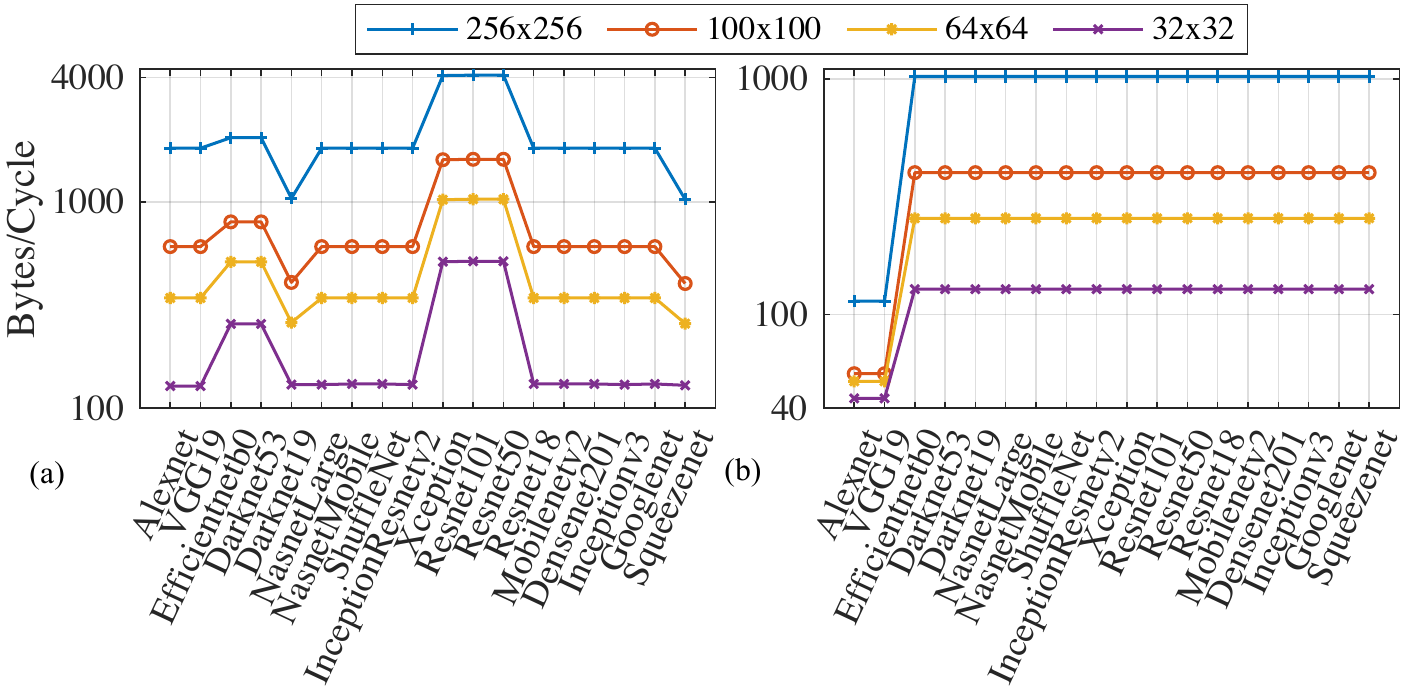}

    \caption{Bandwidth requirement of CV models for different PE array size. (a) Read Bandwidth, (b) Write Bandwidth.}

    \label{bw_cv}
\end{figure}

\begin{figure}[ht]
    \centering
    \includegraphics[scale=0.6]{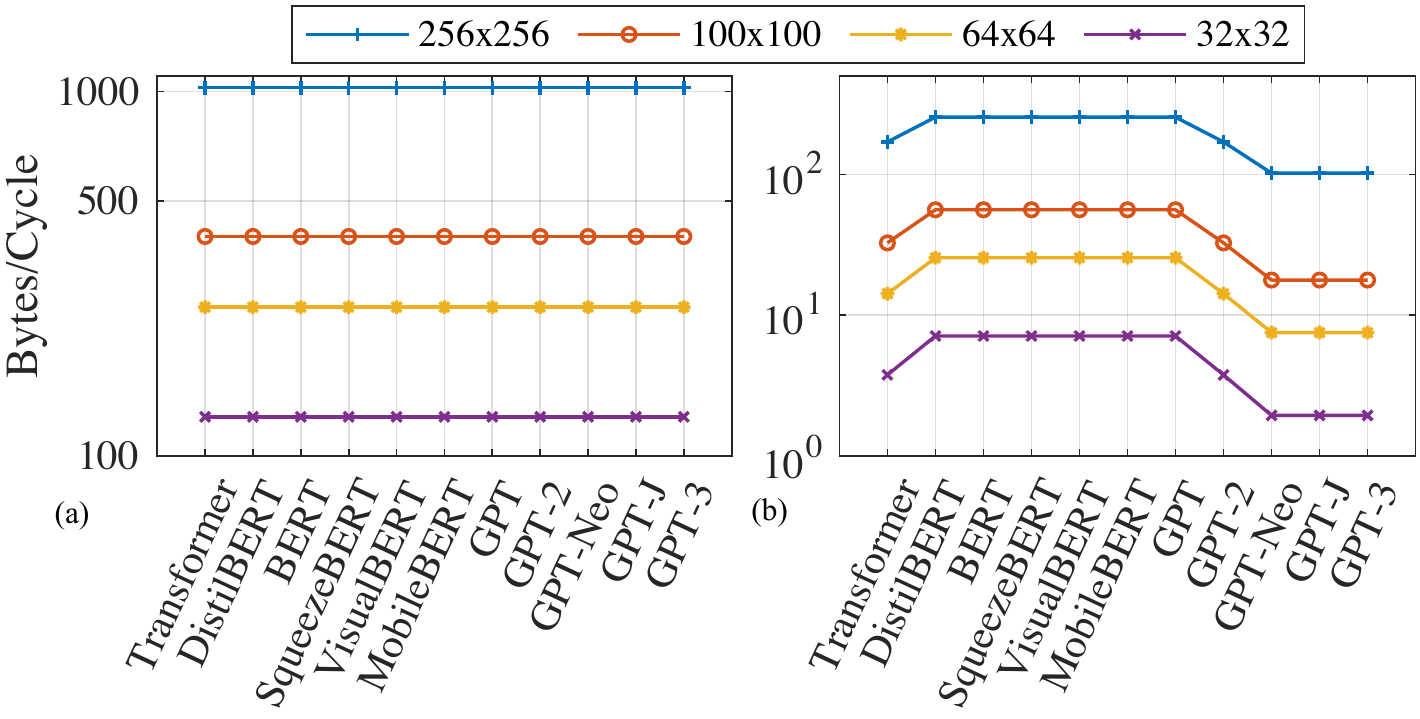}
  
    \caption{Bandwidth requirement of NLP models for different PE array size. (a) Read Bandwidth (for GEMM and softmax operation), (b) Write Bandwidth.}.
  
    \label{bw_nlp}
\end{figure}

\begin{figure*}[ht]
    \centering
    \includegraphics[width = \textwidth]{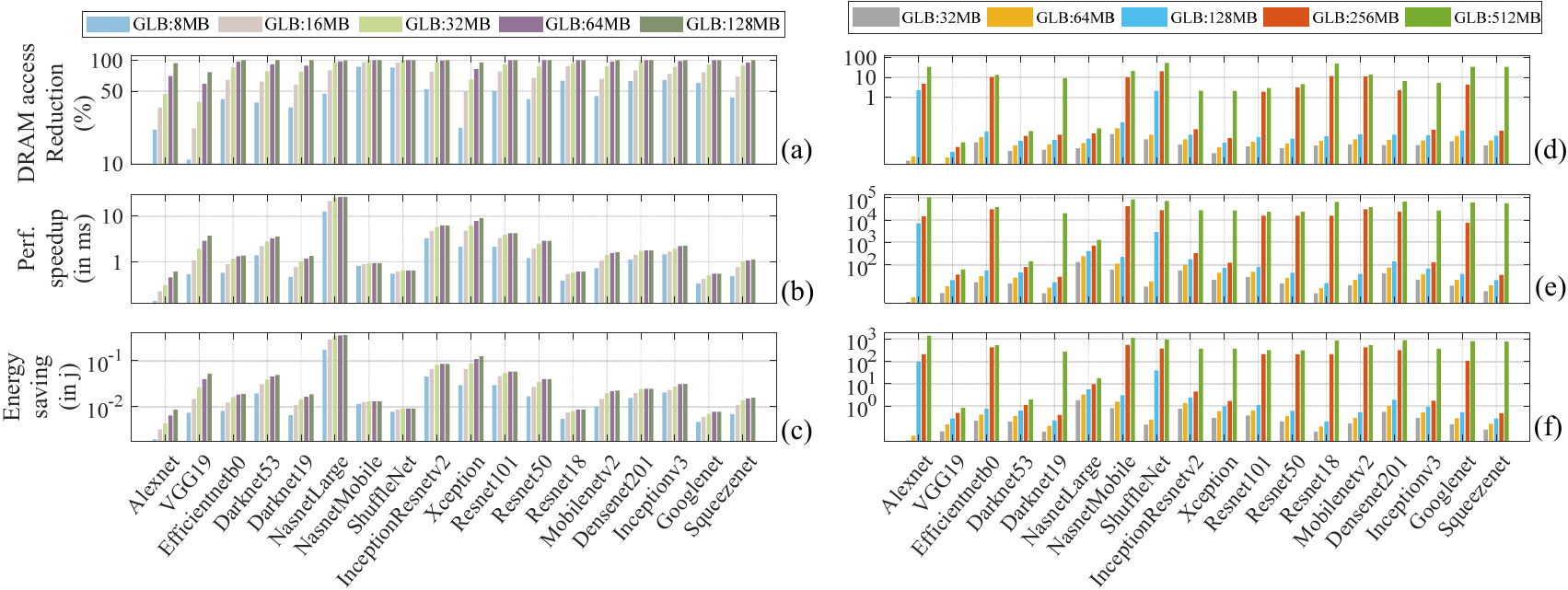}
   
    \caption{Impact of larger GLB memories on performance and energy efficiency for CV models at inference and training. Percentage reduction in DRAM accesses at inference (a) and training (d). Performance Speedup  from DRAM access reductions at inference (b) and training (e). Energy savings from reduced DRAM accesses at inference (c) and training (f). Both cases compare results to a baseline of 2MB GLB running 16 samples.}
   
    \label{d_fig1_cv_infr}
\end{figure*}

\begin{figure*}[ht]
    \centering
    \includegraphics[scale = 1.15]{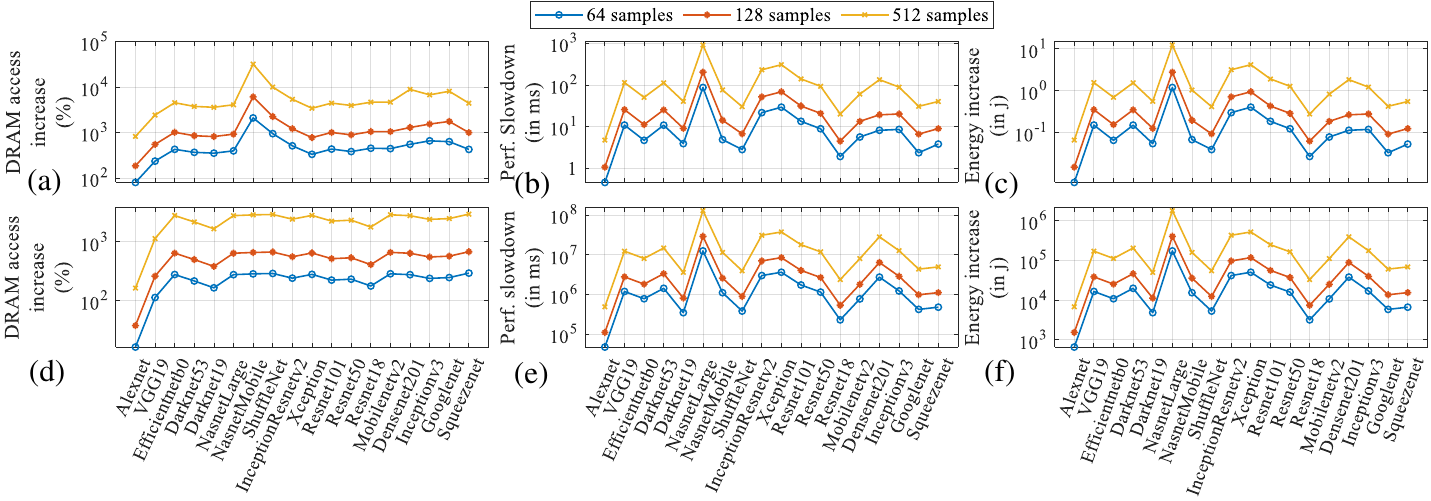}
    
    \caption{Impact of batch size on performance and energy efficiency for CV models at inference and training. Percentage increase in DRAM accesses at inference (a), at training (d). Performance slowdown (latency increase) from extra DRAM accesses at inference (b), at training (e). Energy increase from extra DRAM accesses at inference (c), at training (f). In both cases, results are compared to a baseline of 16 samples running with 4MB GLB.}
    
    \label{d_fig2_cv}
\end{figure*}

In Fig. \ref{bw_cv} (a), (b), we plot the read-write on-chip bandwidth demand in $bytes/cycle$ of 18 widely used CV models. Resenet101 and Resnet50 running on a 256$\times$256 PE array will demand the highest read bandwidth, 4017 bytes/cycle, from GLB, whereas Squeezenet will demand the lowest bandwidth, 1028 bytes/cycle. Naturally, as the PE array size increases, the computation capacity per cycle $T_{MAC}$ increases which demands more data from memory to keep all PEs active. From the workload perspective, we observe that the most contributing factor to the read bandwidth demand is its inverse relationship with the filter and ofmap size. We explain the inverse relationship of filter and ofmap size with the read bandwidth using the convolutional reuse concept. As the filter size decreases, the scope of convolutional reuse decreases. The ofmap again depends on the filter and ifmap size. With the decrease of filter size and ofmap size, the convolutional reuse decreases, giving rise to more bandwidth demand. The layer of Resnet101 that requires the most bandwidth (4017 bytes/cycle) has the ofmap dimension (7$\times$7) and filter dimension (1$\times$1). On the other hand, the most demanding (1028 bytes/cycle) layer of Squeezenet has the ofmap dimension (18$\times$18) and filter dimension (1$\times$1). Another observation is that though 1$\times$1 convolution reduces the computation complexity, it requires more bandwidth from memory, i.e., becomes memory intensive. The write bandwidth is also inversely proportional to the filter size. However, in 1$\times $1 convolutions, it depends on the number of outputs generated by the PE array. The write bandwidth is always smaller than the read bandwidth (Fig. \ref{bw_cv} (b)) as it takes more than one operand to generate one output. For example, in a 3$\times$3 convolution, it takes 18 operands to generate a single output; in a 1$\times$1 convolution, it takes two operands.

\begin{figure*}[ht]
    \centering
    \includegraphics[width = \textwidth]{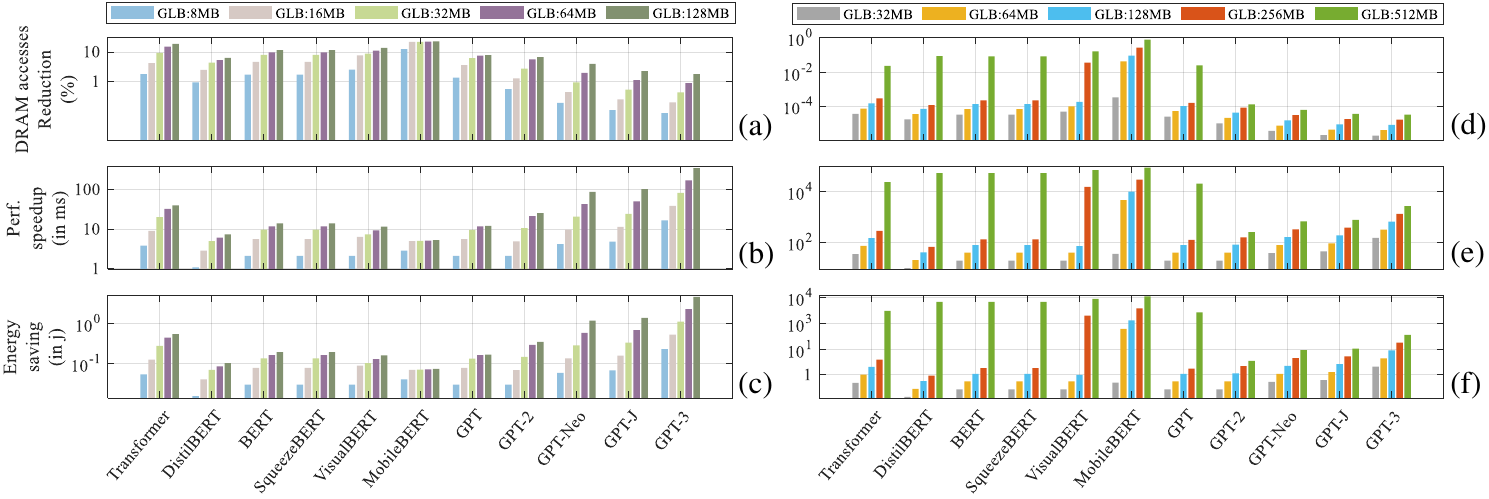}
    
    \caption{Impact of larger GLB memories on performance and energy efficiency for NLP models at inference and training. Percentage reduction in DRAM accesses at inference(a), at training (d). Performance Speedup  from DRAM access reductions at inference (b), at training (e). Energy savings from reduced DRAM accesses at inference (c), at training (e). In both cases, results are compared to a baseline of 2MB GLB running 16 samples}
    \label{d_fig1_nlp}
    
\end{figure*}

\begin{figure*}[ht]
    \centering
    \includegraphics[scale = 1.11]{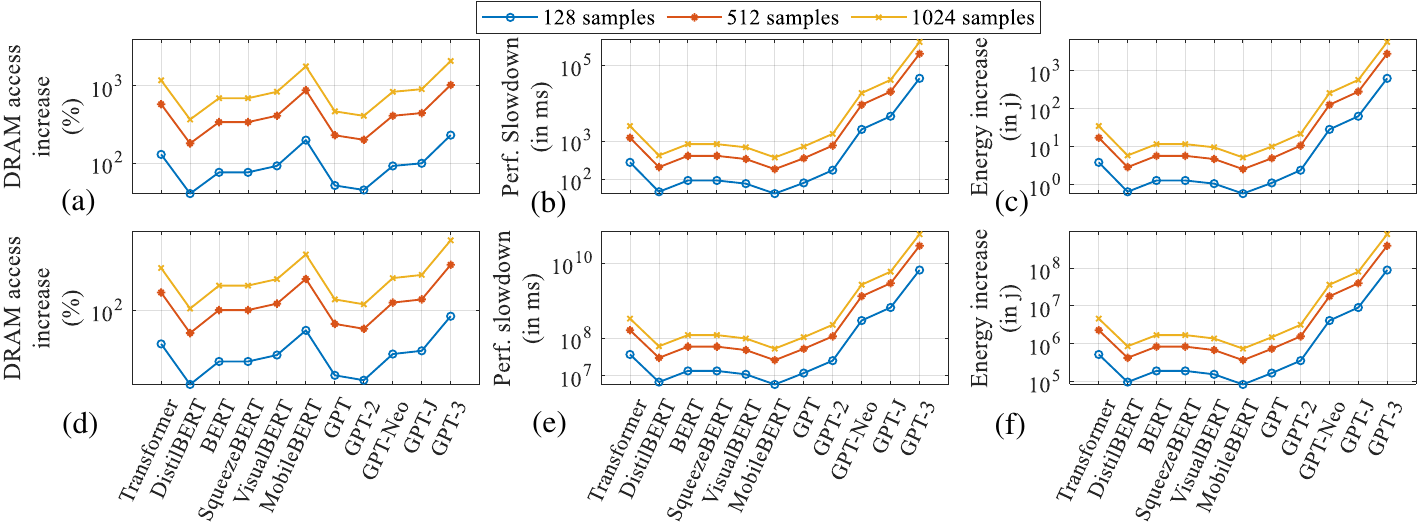}
    
    \caption{Impact of batch size on performance and energy efficiency for NLP models at inference and training. Percentage increase in DRAM accesses, inference (a), and training (d). Performance slowdown (latency increase) from extra DRAM accesses at inference (b), at training (e). Energy increase from extra DRAM accesses at inference (c), at training (f). Results are compared to a baseline of 16 samples running with 4MB GLB.}
    \label{d_fig2_nlp}
\end{figure*}

As mentioned in section \ref{bw_fc_layer}, the bandwidth requirement for transformer-basded model are calculated using the expressions of Table \ref{table:bw_expression}. The dimension of the operand matrices is larger than the PE array dimension, hence following Case IV (Table \ref{table:bw_expression}, Section \ref{bw_fc_layer}), the read bandwidth of all models depends on the PE array size (Fig. \ref{bw_nlp} (a)). The write bandwidth depends on the PE array dimension and the input sequence length. The softmax read bandwidth depends on the SFU width, and matches with the GEMM read bandwidth. 
As different models are trained with different input sequence lengths\cite{hugging_face}, their write bandwidth demand is not the same across all models. The parameter sizes and settings of the models used in this work are shown in Table \ref{nlp_param}.
The models having the highest sequence length (2048) have the lower write bandwidth demand 102 bytes/cycle running on a 256$\times$256 PE array (Fig. \ref{bw_nlp} (b)).

\subsection{Impact of on-chip memory}
\label{on-chip impact}

\begin{figure*}[ht]
    \centering
    \includegraphics[width=\textwidth]{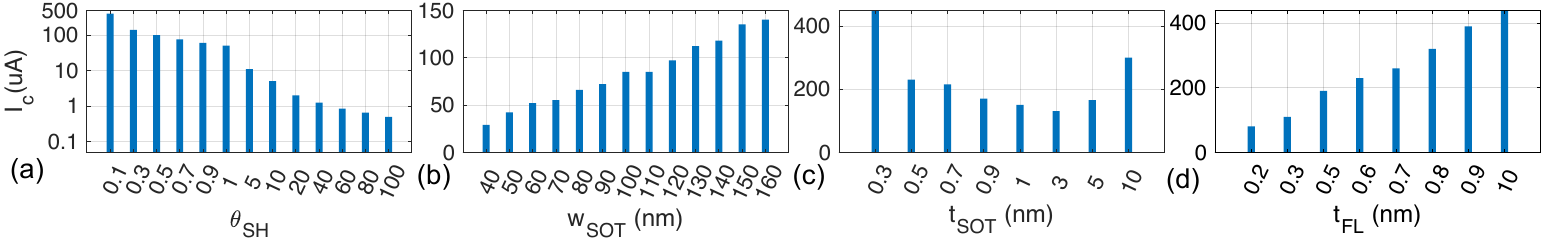}
    
    \caption{Critical current vs $\theta_{SH}$(a), $w_{SOT}$(b), $t_{SOT}$(c), and $t_{FL}$(d).}
    \label{fig:dtco_ic}
    
\end{figure*}


\begin{figure}[ht]
    \centering
    \includegraphics[scale=0.48]{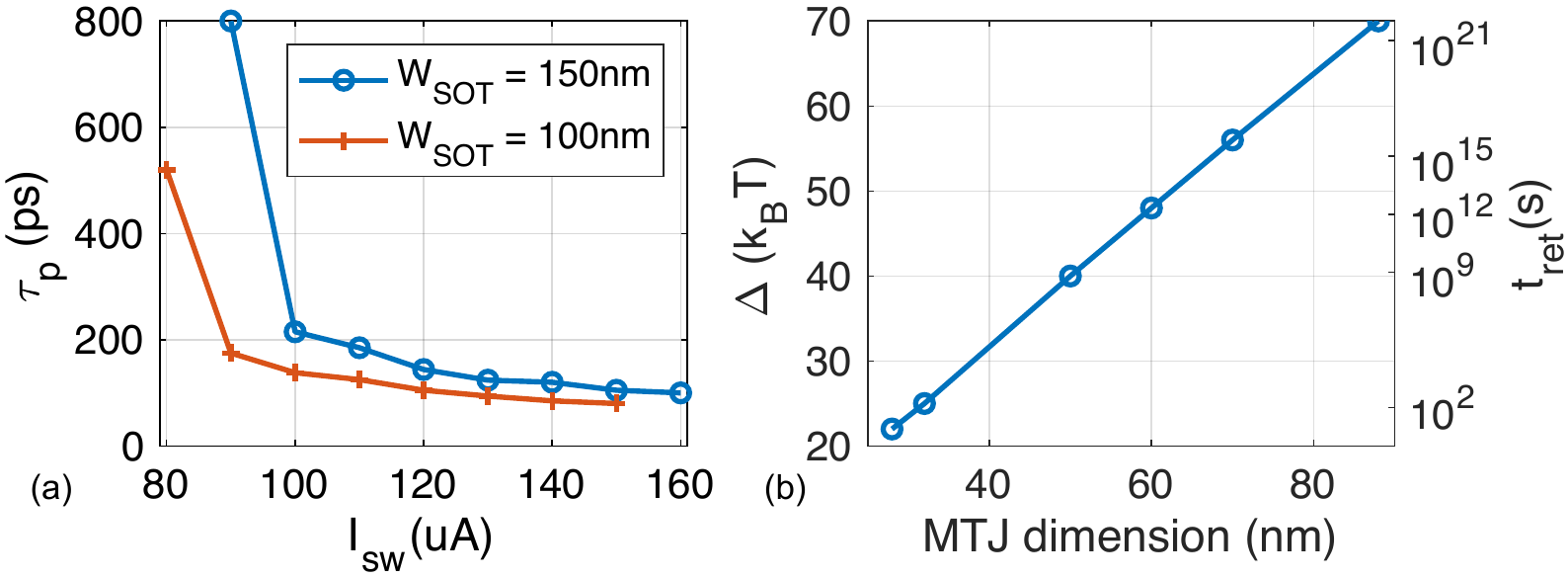}
   
    \caption{(a) Switching pulse width $\tau_{p}$ vs applied switching current $I_{sw}$.  (b) Thermal stability factor $\Delta$ (left Y-axis) and retention time $t_{ret}$ (right Y-axis) vs MTJ dimension for a fixed retention failure rate, $P_{RF} = 10^{-9}$. At $\Delta = 70$, MTJ dimension = 88nm, retention time is $>$ 10 years \cite{imce2019}.}
    \label{fig:dtco_pulse_width}
\end{figure}

\begin{figure}
    \centering
    \includegraphics[scale=0.48]{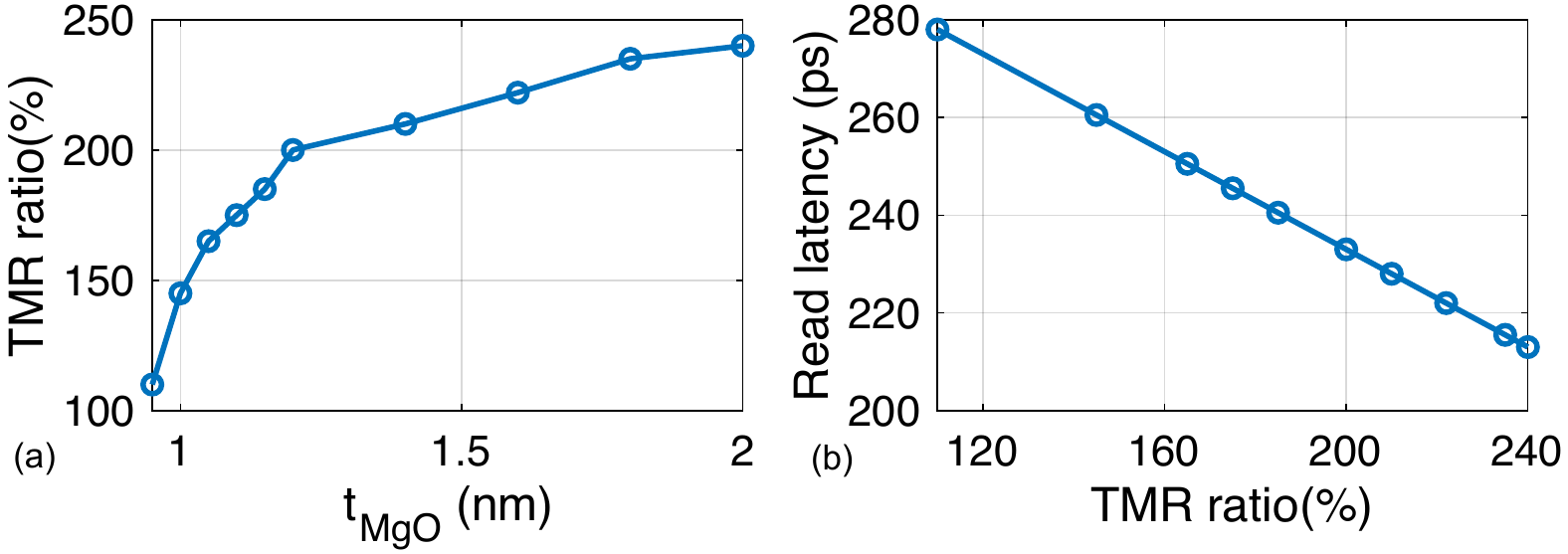}
    
    \caption{Impact of, (a) oxide thickness on TMR. (b) TMR on read latency.}
    \label{fig:tmr_vs_oxide_rd_ltncy}
    
\end{figure}

Compared to a GLB size of 2MB, the DRAM access counts for all CV models decrease significantly if we increase the GLB size. In inference, reaching the 100\% reduction in access means it only needs to read the initial inputs, weights for each layer, and write the final layer output, no DRAM access is needed for the intra and inter-layer operations. Further increase in GLB size will not improve the performance in these cases. 
For 16 samples, DRAM access is reduced by 100\% for 14 models at 128MB, and most models experience a reduction of >80\% at 64MB (Fig. \ref{d_fig1_cv_infr} (a)). Fig. \ref{d_fig1_cv_infr} (b), (c) show the performance speed up and energy saving coming from these DRAM access reductions.

We observe a slower improvement in the DRAM access reduction during training unless the GLB size is large enough, at least 256MB for most models (Fig. \ref{d_fig1_cv_infr} (d)). However, even the smaller percent reduction in DRAM access results in significant performance and energy improvement 
(Fig. \ref{d_fig1_cv_infr} (e), (f)). This is because training requires at least 2$\times$ DRAM accesses as inference. The smaller percent reduction of a large number of DRAM accesses translates to a significant energy and latency improvement. 
A similar trend is observed for NLP models. Transformer-based NLP models are usually larger than the CV models. This is the reason we achieve more performance speedup and energy reduction even at smaller DRAM access reduction rate (Fig. \ref{d_fig1_nlp}).We also observe that DNN models learn faster if we increase the batch size. However, for a fixed GLB size, the DRAM access count increases significantly at larger batch size, causing performance slowdown and more energy consumption. Fig. \ref{d_fig2_cv} and Fig. \ref{d_fig2_nlp} (a, b, c for inference and d, e, f for training) show the increase in DRAM access count and its associated impact on performance and energy for CV and NLP models respectively at different batch sizes.  

The key takeaway from this analysis is that we can reduce the energy and latency associated with DRAM accesses if we increase the GLB size. For larger batch sizes, the energy and latency improvement is even more. Because at large batch sizes, throughput increases at the cost of DRAM accesses. As we increase the GLB size, DRAM accesses reduce, and we achieve latency and energy reduction.

\subsection{DTCO of SOT for PPA Optimization}

From section \ref{on-chip impact} we see that the GLB size of 64MB (for inference) and 256MB (for training) offer significant energy and performance improvement. However, it is not feasible and efficient to use such large SRAMs because of its area and leakage power, even if the low-power techniques are employed. Section \ref{bw_demand} implies that we need approximately 4000bytes/cycle bandwidth between GLB and PE array for larger array size (256$\times$256). In this subsection, we provide the SOT-MRAM DTCO results and observation meeting the requirements stated in the above two subsections. We perform the DTCO in \textit{Cadence Virtuoso}  tool using the compact SOT-MRAM model from \cite{sot_model_kazemi}, and use \textit{Synopsys} 14nm library \cite{saed14} for  the CMOS transistors and peripheral circuits.

\subsubsection{$I_{C}$ optimization}

To realize the impact of SOT efficiency $\theta_{SH}$ on $I_{c}$, we sweep $\theta_{SH}$ from 0.1 to 100 (Fig. \ref{fig:dtco_ic} (a)). With $\theta_{SH} \ge 100$,  $I_{c}$ goes as low as 0.5uA. Even though the widely used SOT layers are made of heavy metal alloys having smaller $\theta_{SH}$ (e.g., 0.1 to 0.5), recent advancement in material engineering demonstrates that in topological insulator $\theta_{SH}$ can go as high as 152 \cite{khang2018conductive}. We recommend using topological insulators as the SOT layer to achieve a lower switching current.

Next, we analyze the impact of SOT layer geometry on the switching current (Fig. \ref{fig:dtco_ic} (b), (c)). $I_{c}$ scales down linearly with the decrease of SOT layer width, and $w_{SOT}$ can be set to desired value based on the performance and reliability requirement (Fig. \ref{fig:dtco_ic} (b)). While $I_{c}$ scales linearly with the width of the SOT layer, the thickness of the SOT layer has an interesting effect on the switching current. The SOT layer should be relatively thin but bulk enough for heavy metal layers to experience the bulk effect to achieve high SOT efficiency. Once it crosses optimum thickness, which is ~3nm (Fig. \ref{fig:dtco_ic} (c)), many of the charges that are injected into the metal do not contribute to the switching, and $I_{c}$ increases. 

The smaller the free layer thickness, $t_{FL}$, the smaller the switching current (Fig. \ref{fig:dtco_ic} (d)). We also scale the diameter of MTJ, $d_{MTJ}$, to reduce the MTJ area. However, with the scaling down of $d_{MTJ}$ together with $t_{FL}$, the thermal stability factor $\Delta$ also scales down, reducing the memory's data retention time $t_{ret}$. Non-volatility is a great feature of MRAM, but it can be compromised to achieve higher density, higher bandwidth, and lower energy when the target application is a cache. Because, in the cache even for AI workloads, the data lifetime is much shorter, typically in the seconds range \cite{stt_ai_us}. Fig. \ref{fig:dtco_pulse_width}(b) shows $\Delta$ and $t_{ret}$ as functions of free layer volume. While scaling down $t_{FL}$ to optimize $I_{c}$, and $d_{MTJ}$ to optimize area, we keep an eye on the reliability of the stored data. We consider a retention failure rate of $10^{-9}$ (i.e., 1 bit flip per billion).

\subsubsection{Bandwidth optimization}
As shown in Fig. \ref{fig:tmr_vs_oxide_rd_ltncy} (a), TMR ratio of the MTJ device can be increased by increasing the oxide thickness \cite{tsunekawa2005giant}. We increase the oxide thickness to decrease the  read latency (Fig. \ref{fig:tmr_vs_oxide_rd_ltncy} (b)). 
The write pulse width is inversely proportional to the applied switching current. While we want to lower the applied current to achieve low energy, the higher amplitude of the applied current is required for faster magnetization reversal. However, switching occurs at smaller pulse width at the iso-current if we scale down the SOT layer width. This is because of the smaller critical current at smaller geometry (Fig. \ref{fig:dtco_ic} (b,d)).
Fig. \ref{fig:dtco_pulse_width}(a) shows that switching pulse width can be reduced significantly by scaling down the SOT layer width. Thus, we can achieve higher write bandwidth by scaling down the SOT layer width to meet the high BW demand from AI workloads. 


\begin{table}[ht]
\centering
\caption{SOT-MRAM DTCO optimized parameters. 30\% guard-band are added with thickness and width for process variations.}
\label{optimized_dtco_param}

\setlength\tabcolsep{5pt}
\begin{tabular}{|l|l|l|l|}
\hline
\multicolumn{1}{|c|}{Parameter} & \multicolumn{1}{c|}{Value} & \multicolumn{1}{c|}{Parameter} & \multicolumn{1}{c|}{Value} \\ \hline
Spin Hall angle $\theta_{SH}$   & 1                          & TMR                            & 240\%                      \\ \hline
Free layer thickness $t_{FL}$   & 0.5nm                      & MTJ diameter $d_{MTJ}$               & 55nm                       \\ \hline
SOT width $w_{SOT}$             & 130nm                      & SOT thickness $t_{SOT}$        & 3nm                        \\ \hline
Oxide thickness $t_{MgO}$       & 3nm                        & Thermal stability factor $\Delta$                       & 45                         \\ \hline
\end{tabular}

\end{table}

\subsection{Process \& Temperature Variation and Bitcell Simulation}
In this subsection, we perform Process and Temperature variation on the DTCO-optimized parameters, design the peripheral circuits, and test the read-write operation on the bit cell at scaled parameters.

\subsubsection{Process and Temperature variation}
\begin{figure}[h]
    \centering
\includegraphics[width=0.43\textwidth]{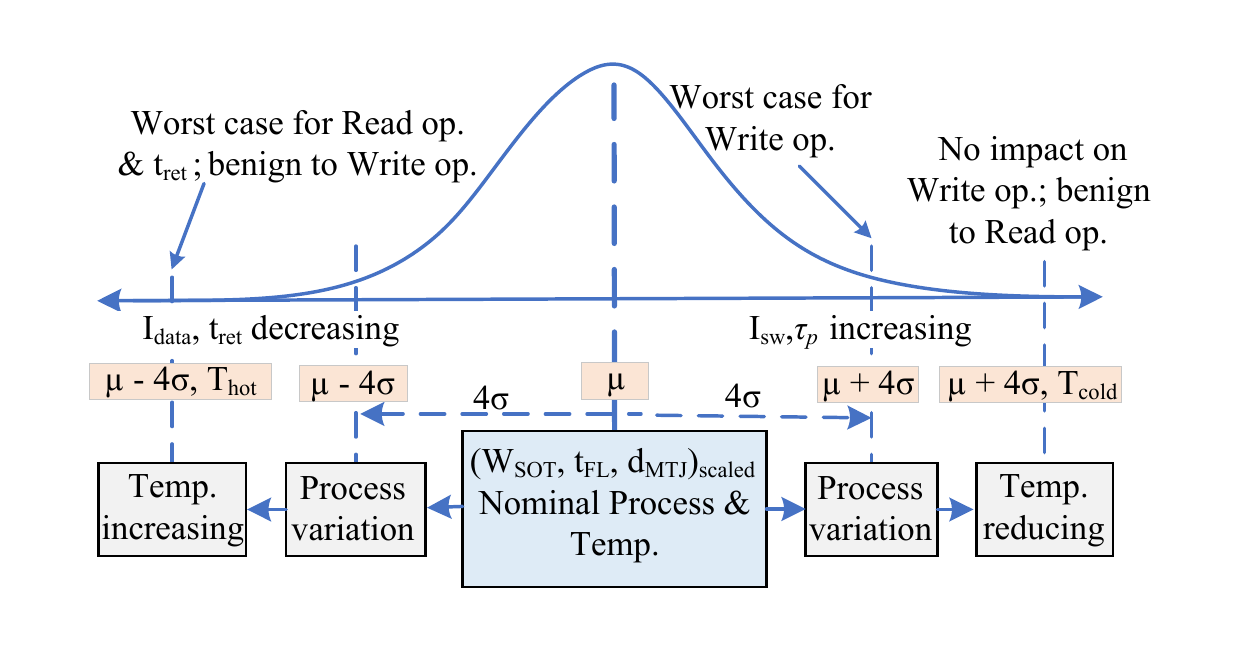}
    \caption{Impact and distribution of Process and Temperature variation on scaled parameters.}
    \label{fig:process_var}
\end{figure}

To incorporate process variations, we model MTJ diameter, free layer thickness, and SOT layer width as Gaussian variables in the Verilog A model of SOT-MTJ \cite{sot_model_kazemi}. We assume standard deviations ($\sigma$) as 5\% of their respective means ($\mu$) and perform Monte Carlo simulations with 5000 samples within 4$\sigma$ variation. We also consider the temperature variations. The extreme point at the right side of the scaled target parameter is $\mu + 4\sigma, T_{cold}$ (Fig. \ref{fig:process_var}). From equations \ref{Ic3} and \ref{tau}, $I_{sw}$ and $\tau_{p}$ are independent of Temperature. As a result, the worst case for write operation (highest $I_{sw}$ and longest $\tau_{p}$) is at $\mu + 4 \sigma$. This point is, however, benign to the read operation and retention failure.
As we scale down $d_{MTJ}$ and $t_{FL}$, $\Delta$ also reduces, reducing $t_{ret}$, and $I_{data}$. $\Delta$ reduces further as temperature increases \cite{stt_eqn1}. Thus, the worst case for read operation (smallest $I_{data}$) and retention failure (smallest $t_{ret}$) is at $\mu - 4 \sigma, T_{hot}$ (see Fig. \ref{fig:process_var}). As $I_{data}$ reduces, the difference between $I_{data1}$ and $I_{data0}$ becomes even smaller and difficult to sense.

To ensure the reliability of the SOT-MRAM bit cell, we add a 30\% guard band on the scaled SOT device parameters: 20\% for process variation and 10\% for temperature variation. The optimized DTCO parameters after adding the PT induced 30\% guard-band are shown in Table \ref{optimized_dtco_param}.

\subsubsection{Write operation}
To write SOT-MRAM bitcell, we bias BL with the data-to-be-written and SL with the complement of data-to-be-written. Assuming that the magnetization state of the Reference layer is -1, to write 1 into the MTJ bitcell, we switch the magnetic orientation of the Free layer to +1 state resulting in a high resistive state. To achieve this state, we turn on the WWL, connect BL to VDD and SL to the ground. The resultant current switches the free layer's magnetic orientation from -1 to +1. The opposite bias is applied to write 0. We do not need any additional peripheral circuits for the write operation of SOT-MTJ.

\subsubsection{Read operation} 
\begin{figure}
    \centering    \includegraphics[scale=0.43]{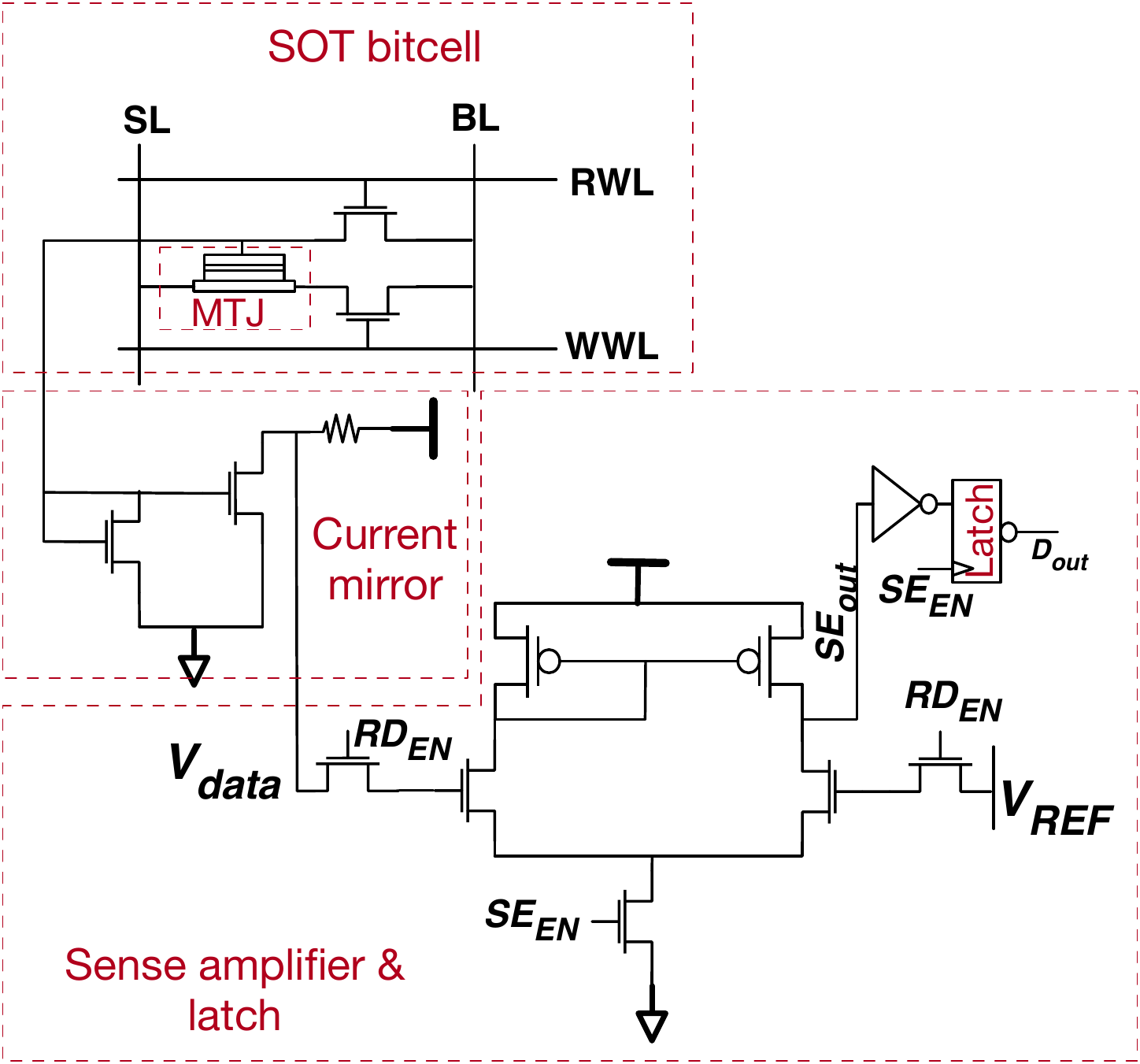}
    \caption{SOT-MTJ bitcell with read sensing circuitry.}
\label{fig:peripheral_circuits}
\end{figure}

Read operation involves sensing the current passing through MTJ at P and AP states. For our SOT-MRAM bitcell, with the parameters shown in Table \ref{optimized_dtco_param}, $I_{data0} = 20uA$ and $I_{data1} = 33uA$. We design and optimize the read circuitry to sense this small differential current, as shown in Fig. \ref{fig:peripheral_circuits}. 
Our proposed read sensing circuit only contains an additional current mirror block (to amplify current), and it does not require the precharge circuits compared to SRAM. Hence, there is no additional area overhead in the periphery compared to SRAM. The dynamic power consumption are shown in Table \ref{tab:power}

To capture the stochastic nature of MTJ switching, we simulate the bit cell for 1000 bitstream. We achieve a read and write yield of 100\%, and at 250ps and 520ps, respectively.  
This results in  read bandwidth of 4 Gbps and a write bandwidth of 1.9 Gbps. We then dynamically allocate the memory bus width on-demand to satisfy the bandwidth requirement for different workloads and PE array size stated in section \ref{bw_demand}.
\begin{table}[h]
    \centering
    \caption{Dynamic Power consumption (in uW) of SRAM and SOT-MRAM. (1/0) means the corresponding power to access bit 1 and 0.}
    \begin{tabular}{|c|c|c|}
        \hline
         & Read(1/0) & Write(1/0) \\ \hline
        SRAM & 426 & 373 \\ \hline
        SOT-MRAM & 150/368 & 325/300 \\ \hline
        \end{tabular}
    \label{tab:power}
\end{table}

\subsection{System level performance evaluation of SOT-MRAM based Memory}
\begin{figure*}[ht]
    \centering
    \includegraphics[width = \textwidth]{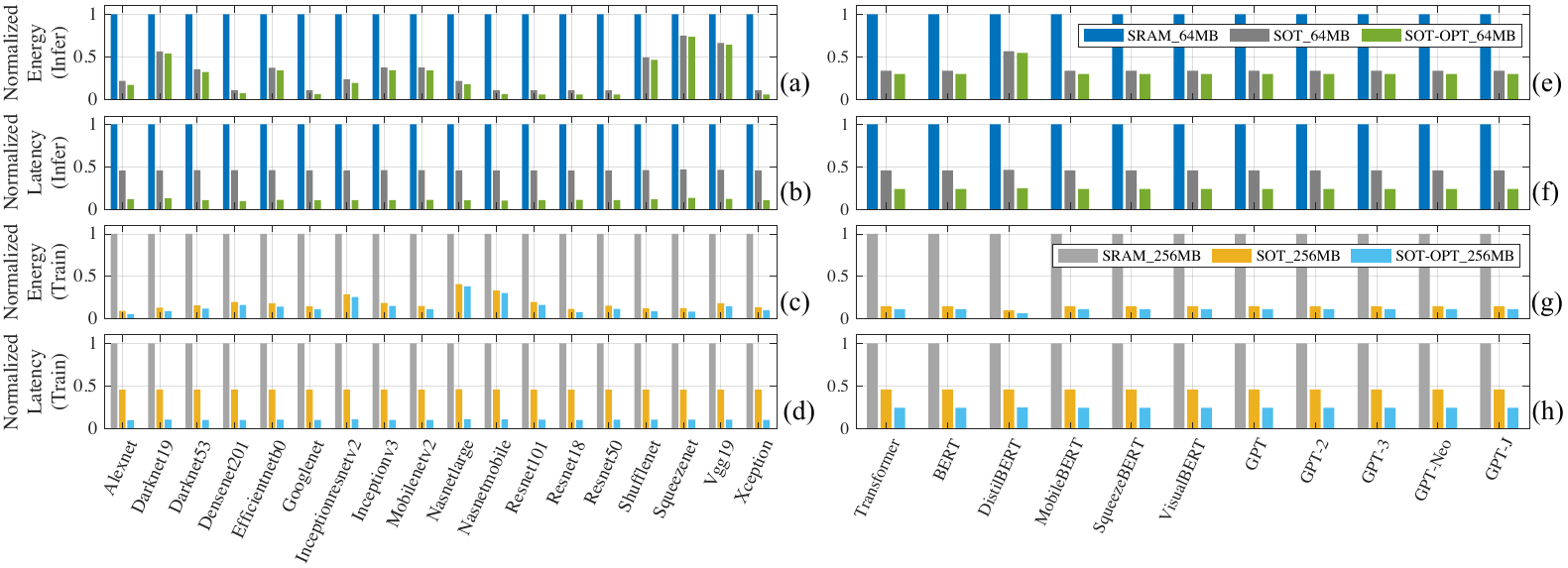}

    \caption{System level energy improvement with SOT-MRAM and DTCO-optimized-SOT-MRAM over SRAM at the same size for CV (a-d) and NLP (e-h) models. The top plots show energy (a, e) and latency (b, f) for inference, and the bottom plots show energy (c, g) and latency (d,h) for training.}

    \label{fig_sys_lvl_perf_cv}
\end{figure*}



In this subsection, we analyze the PPA (Power, Performance, and Area) metrics at the system level on the DNN/CNN benchmarks with SRAM, SOT-MRAM, and DTCO-optimized-SOT-MRAM. 
We use the Destiny \cite{destiny} memory simulator to find the array-level data for both SRAM and SOT-MRAM. We modify Destiny source code to reflect: (i) SOT switching mechanism, (ii) special read sensing circuit for SOT-MRAM, and (iii) 14nm CMOS technode. Then, we feed the extracted bitcell-level data of SOT-MRAM in the $.cell$ file to find the PPA at the desired memory capacity.

\begin{figure}[ht]
    \centering
    \includegraphics[scale = 0.55]{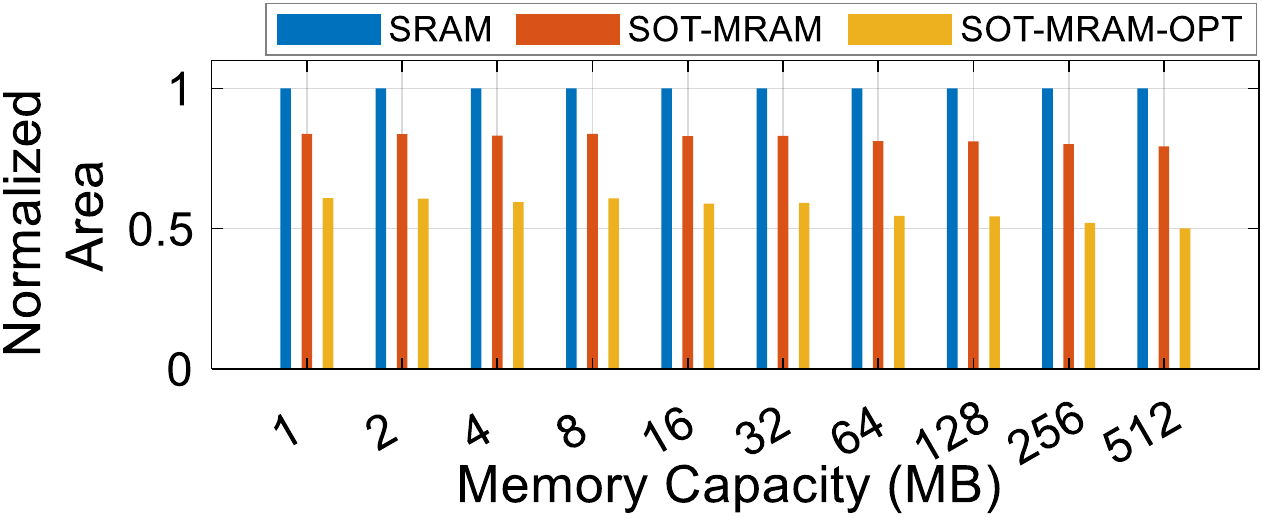}

    \caption{Area improvement of SOT-MRAM and SOT-MRAM-OPT}
    \label{sys_area_comp}
   
\end{figure}

 Based on the array-level results from Destiny, and DRAM \& GLB access counts from Algorithms \ref{alg:dram_acc_cnt_infr}, and \ref{alg:dram_acc_cnt_trng}, we estimate the system-level power and performance. Finally, we analyze the area of the memory modules of different technologies (14nm SRAM, SOT-MRAM, and DTCO-opt-SOT-MRAM) at iso-capacity. This analysis only incorporates the PPA metrics from the memory system (DRAM and GLB), assuming that the PPA of the compute unit is constant. With SOT-MRAM as GLB, we see significant energy and latency improvement over SRAM at 64MB (for inference) and 256MB (for training) (see Fig. \ref{fig_sys_lvl_perf_cv} (a-d) for DNN benchmarks and (e-h) for NLP benchmarks). On average, the 64MB SOT-MRAM offers 5$\times$ energy reduction and 2$\times$ latency reduction over 64MB SRAM across all CNN models at inference. Our DTCO-optimized-SOT-MRAM offers further improvement, 7$\times$ energy, and 8$\times$ latency reduction over SRAM at iso-capacity. For latency improvement, the most contributing factor is the DRAM access reduction with large GLB and the smaller read/write latency of SOT-MRAM at larger capacity compared to SRAM. At smaller capacity, SRAM is way faster than SOT-MRAM \cite{tahoori_1,optimized_SOT_imec}. We observe that the most contributing factor in energy reduction (>50\%) is the near-zero leakage power of SOT-MRAM compared to high leakage power of SRAM. The improvement is even more in training mode; 6$\times$ (8$\times$ with SOT-opt.) energy reduction and 2$\times$ (9$\times$ with SOT-opt.) latency reduction. With 64MB SOT-MRAM, NLP models in inference mode experience 2$\times$ (3$\times$ with SOT-opt.) energy reduction and 2$\times$ (4$\times$ with SOT-opt.) latency reduction than 64MB SRAM. Like CV benchmarks, with 256MB SOT-MRAM, NLP benchmarks also experience more energy improvement, 6$\times$ (8$\times$ with SOT-opt.), and latency improvement, 2.5$\times$ (4.5$\times$ with SOT-opt.), in training mode. The more improvement in training mode is because of two reasons: (1) GLB size increases from 64MB to 256MB, and (ii) GLB access counts are significantly large (at least 5$\times$) in training. Our DTCO-opt-SOT-MRAM further adds value to PPA by its smaller silicon area, 0.54$\times$ at 64MB and 0.52$\times$ at 256MB of 14nm SRAM at iso-capacity (Fig. \ref{sys_area_comp}).

\section{Related Work}
\label{rltd_work}
SOT-MRAMs have been widely studied as the next generation of STT-MRAM to leverage all benefits of MRAMs as embedded memory \cite{recent-progress-in-SOT_fab3}\cite{optimized_SOT_imec}\cite{dualport_fieldfree_fab2}\cite{ultrafast_embedded_mem_fab4}\cite{size_dependent_switching_fab1}\cite{sot_0.35ns_write}. However, very few studies have evaluated the performance of SOT-MRAM as on-chip memory in system-level for AI accelerators. \cite{tahoori_1} and \cite{sys_lvl_eval_sot_rltd_wrk} demonstrated the performance improvement of SOT-MRAM as L2 data cache compared to SRAM L2 cache on MiBench, SPEC2000 and SPEC2006 benchmarks. SOT-MRAMs have also been explored in the context of DL accelerators as a promising technology for In-Memory Computing (IMC) or Computing-In Memory (CIM) \cite{PXNOR_BNN} \cite{CMP_PIM} \cite{IMC1_for_RW}\cite{IMC2_for_RW}\cite{IMC4_for_RW}. IMC/CIM over conventional AI accelerator has pros and cons, and the detailed comparison between these two domains is outside the scope of this work.  Our work, where we use SOT-MRAM as the cache storage element, differs from crossbar-based in-memory computing.
While the scope of SOT-MRAM has been explored both as regular CPU cache and IMC for DL accelerator to some extent, to the best of our knowledge, unlike IMC, 
this is the first work that presents a comprehensive analysis of SOT-MRAM as on-chip memory for application in AI/DL accelerators.

\section{Conclusion}
\label{conclusion}
In this research, we presented a System and  Technology Co-optimization methodology for efficient and high-performance memory system design with SOT-MRAM for modern AI accelerators. Guided by detailed target workload characterization, our memory system comprises of HBM3 DRAM, a DTCO-enabled SOT-MRAM GLB and a small SRAM buffer. Our large SOT-MRAM GLB significantly reduces the energy and latency by reducing expensive DRAM accesses while still having acceptable on-chip access energy and latency, achieving overall system-level high performance. We finally demonstrate that our memory system performs 8$\times$ and 9$\times$ better in terms of energy and latency respectively on CV benchmarks in training (7 and 8 times better in inference) and 8$\times$ and 4.5$\times$ better in terms of energy and latency respectively on NLP benchmarks in training (3 and 4 times better in inference) while consuming only around 50\% of SRAM area at iso-capacity.


%





\ifCLASSOPTIONcaptionsoff
  \newpage
\fi



\bibliographystyle{IEEEtran}

%
\bibliography{REFERENCE_NEW.bib}


%

\begin{IEEEbiography}[{\includegraphics[width=1\linewidth]{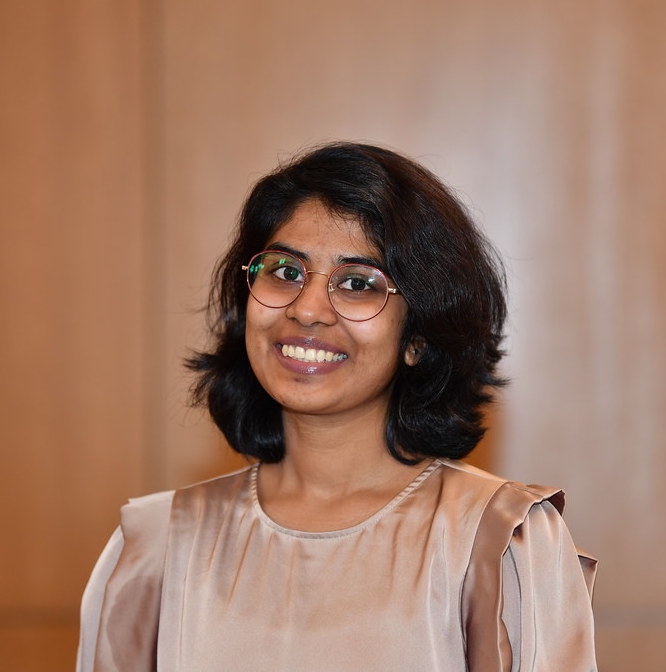}}]{Kaniz Mishty} received the B.S. degree in Electronics and Communication Engineering from Khulna University of Engineering and Technology, Bangladesh, in 2018. She is currently working towards her Ph.D. degree in ECE at Auburn University, AL, USA. Her research interests are energy efficient AI hardware design and AI/ML in CAD. She interned with Apple Inc. in Summer '22, 23 and Qualcomm in '21   on AI application in SoC and custom circuit design.
\end{IEEEbiography}

\begin{IEEEbiography}[{\includegraphics[width=\linewidth]{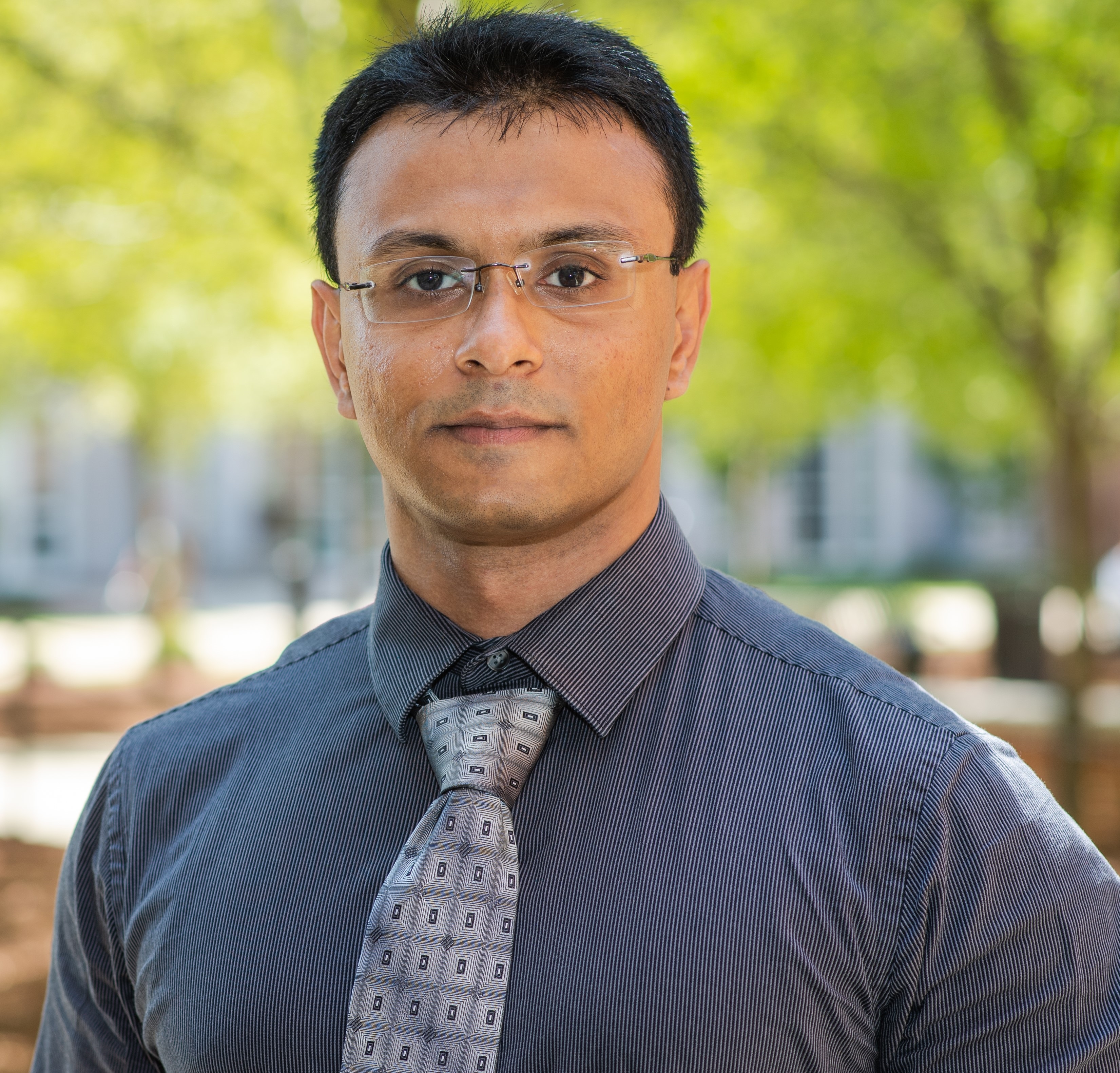}}]{Mehdi Sadi} (S'12-M'17) is an Assistant Professor at the Department of Electrical and Computer Engineering at Auburn University, Auburn, AL.  Dr. Sadi  earned his PhD in ECE from  University of Florida, Gainesville, in 2017, MS from University of California at Riverside, USA in 2011 and BS from Bangladesh University of Engineering and Technology in 2010.   Prior to joining Auburn University, he was a Senior R\&D SoC Design Engineer at Intel Corporation in Oregon. Dr. Sadi`s research focus is on developing algorithms/CAD techniques for implementation, design, reliability, and security of AI hardware. He was the recipient of SRC best in session award, Intel Xeon Design Group recognition awards, and National Science Foundation CRII award.
\end{IEEEbiography}








\end{document}